\documentclass[12pt,letterpaper]{article}

\usepackage{graphicx}
\usepackage{xcolor}
\usepackage{tikz}
\usepackage{comment}
\usepackage{amssymb}
\usepackage{amsmath}
\usepackage{amsfonts}
\usepackage{booktabs}
\usepackage{eurosym}
\usepackage{geometry}
\usepackage{ulem}
\usepackage{setspace}
\usepackage{sectsty}
\usepackage{footmisc}
\usepackage{caption}
\usepackage{natbib}
\usepackage{pdflscape}
\usepackage{subfigure}
\usepackage{array}
\usepackage{enumerate}
\usepackage{bm}
\usepackage{url}
\usepackage{listings}
\usepackage{breqn}
\usepackage{braket}
\usepackage{mathrsfs}
\usepackage{mathtools}
\usepackage{here}
\usepackage{bbm}

\usepackage{hyperref}

\mathtoolsset{showonlyrefs=true}

\newtheorem{thm}{Theorem}
\newtheorem{cor}{Corollary}
\newtheorem{prop}{Proposition}
\newenvironment{proof}[1][Proof]{\noindent\textbf{#1.} }{\ \rule{0.5em}{0.5em}}

\newtheorem{lem}{Lemma}
\newtheorem{ex}{Example}

\newtheorem{dfn}{Definition}

\newcommand{\ctext}[1]{\raise0.2ex\hbox{\textcircled{\scriptsize{#1}}}}

\hypersetup{
  colorlinks=true,
  citecolor=blue,
  linkcolor=red,
  urlcolor=orange,
}

\begin{document}

  \bibliographystyle{ecta}

  \title{Dynamic Non-Bayesian Persuasion}
  \author{Masanori Kobayashi\thanks{
Department of Economics, University of California, Los Angeles.
Email: masakobayashi@g.ucla.edu.
I thank Yaron Azrieli, Alexander Bloedel, Simon Board, Daniel Clark, Rachana Das, Jay Lu, Moritz Meyer-ter-Vehn, Ichiro Obara, Mark Whitmeyer, Frank Yang, and Kai Hao Yang for many helpful comments.
I also appreciate feedback from participants at the 2026 Southwest Economic Theory Conference.
}}
  \date{\today}
  \maketitle
  \begin{abstract}
  If a sender in a persuasion game can use a sequence of experiments rather than a single experiment, does this change the sender's value? We show, within a large class of non-Bayesian updating rules, divisibility, introduced in \cite{cripps2018divisible}, exactly characterizes the receiver's updating rules under which the sender is indifferent between static and dynamic persuasion in any environment. Consequently, restricting attention to static persuasion is without loss precisely under divisible updating rules.
  \end{abstract}

\section{Introduction}
  
The provision of strategically designed information, or persuasion, is pervasive in economic interactions. Modeling persuasion in a way that accommodates realistic belief updating is important but challenging, partly because receivers may deviate from Bayesian updating. A growing empirical literature documents such biases. Field evidence points to non-Bayesian updating in professional decision-making, for example in \cite{bhuller2025feedback} and \cite{bordalo2020overreaction}. Laboratory evidence also documents deviations such as underreaction and base-rate neglect \citep{benjamin2019errors}. Biases may also reflect misspecification of the data-generating process: as \cite{bohren2023behavioral} show, many non-Bayesian updating rules are observationally equivalent to Bayesian updating under misspecification.
  
Such non-Bayesian updating may require explicitly modeling the timing of information provision. In many real-world applications, information is provided sequentially rather than all at once. When an agent is non-Bayesian, sequential and simultaneous signals can lead to different posterior beliefs, whereas Bayesian posteriors are invariant to the timing of information provision \citep{cripps2018divisible}. Thus, the sender may use timing as an additional persuasion instrument, potentially increasing the value of persuasion.

Dynamic information provision can be analytically complicated. An analyst may therefore want to balance tractability with realism about biases in the receiver's belief updating. This raises a natural question: under which non-Bayesian updating rules is a static persuasion model, which is more tractable than a dynamic one, without loss?

This paper characterizes such rules within a restricted but still rich class of updating rules. The characterization is based on \textit{divisibility}, a property introduced by \cite{cripps2018divisible}. Divisibility requires that processing multiple signals sequentially or simultaneously lead to the same posterior belief. We show that this invariance extends to the sender's value of persuasion.

To establish this result, the paper studies a dynamic persuasion model in which a sender uses a sequence of experiments rather than a single experiment. The receiver may be non-Bayesian: we allow for a broad class of updating rules, \textit{systematic distortions}, introduced by \cite{de2022non}. Under such distortions, the receiver forms her posterior as if she first computes Bayesian posteriors from signals and then distorts them using a function that depends on the prior. This class nests many widely used non-Bayesian rules, including Grether's $\alpha$--$\beta$ rule \citep{grether1980bayes,benjamin2019errors}, in which an agent geometrically distorts the base rate and the likelihood of signal realizations. The sender commits to a dynamic persuasion strategy, and the receiver updates after each step. Once the final update is complete, the receiver takes an action. 

We begin with the judge-prosecutor example from \cite{kamenica2011bayesian}, where the receiver updates according to Grether's $\alpha$--$\beta$ rule. When the receiver discounts information from the prior, $\alpha \in (0,1)$, the sender can obtain a higher payoff under dynamic persuasion than under static persuasion. Thus, static persuasion is not without loss (Proposition \ref{prop:grether}).

Our main result (Theorem \ref{thm:divisible_iff_indifference}) characterizes when static non-Bayesian persuasion is without loss in any persuasion environment. Under mild regularity conditions, divisible updating rules imply that the sender's value is the same under static and dynamic persuasion in every environment. Conversely, if an updating rule is not divisible, there exists an environment in which the sender strictly prefers either static or dynamic persuasion, so timing is payoff-relevant. Grether's $\alpha$--$\beta$ rule with $\alpha = 1$, which we call \textit{geometric distortion}, is divisible. Bayes' rule is the special case with $\beta = 1$. 

Our contribution is to identify exactly when the tractable static model of non-Bayesian persuasion remains valid in dynamic information environments. Under divisible updating rules, the sender can use the concavification approach of \cite{de2022non} without loss, even when information can be delivered sequentially. When divisibility fails, timing itself can become payoff-relevant.

Although the assumptions behind Theorem \ref{thm:divisible_iff_indifference} are natural, one of them, \textit{prior independence under certainty}, excludes some commonly used non-Bayesian updating rules. This axiom requires that, after a fully revealing signal, the posterior not depend on the prior. One example is linear underreaction, under which the agent's posterior is a convex combination of the Bayesian posterior and the prior. Theorem \ref{thm:weakly_divisibile_iff_indifference} therefore considers an alternative requirement: an uninformative signal should leave beliefs unchanged, which linear underreaction satisfies. Under this condition, we obtain a similar characterization for environments with state-independent sender preferences. The characterizing property is a new condition, \textit{weak divisibility}, that we define for this purpose. It allows deviations from divisibility up to an affine transformation. Since linear underreaction is not weakly divisible, static persuasion is not without loss: in some persuasion environment, the sender strictly prefers dynamic persuasion.

\textit{Related Literature}. 
Our paper relates to two literatures: static and dynamic persuasion with flexible information provision, and deviations from Bayesian updating.

The static Bayesian persuasion framework was introduced by \cite{rayo2010optimal} and \cite{kamenica2011bayesian}. Existing dynamic extensions, such as \cite{che2023keeping}, often allow the receiver to choose after each signal realization whether to act or wait for more information. In our model, by contrast, the receiver acts only after the entire sequence of signal realizations. This isolates the effect of the timing of information provision from the timing of action.

Within the literature on non-Bayesian updating, our paper is closely related to work on systematic distortions, in which updated beliefs are distorted Bayesian posteriors. Examples include Grether's $\alpha$--$\beta$ rule \citep{grether1980bayes,benjamin2019errors}, linear distortion \citep{epstein2010non}, and, more recently, two-stage updating under cognitive constraints \citep{ba2025over}. The most relevant paper is \cite{cripps2018divisible}, which introduces divisible updating rules and characterizes their functional form under several axioms. Our contribution is to establish an implication of divisibility for dynamic persuasion.
A related but distinct property of belief distortions is \textit{Gretherian coherency}, introduced by \cite{chambers2023coherent}. A distorted belief is Gretherian-coherent if the belief obtained by first computing the Bayesian posterior and then applying a distortion function coincides with the belief obtained by first applying the same distortion function to the prior and then Bayesian-updating it with a suitably distorted experiment. While divisibility requires equivalence between simultaneous and sequential updating from multiple signals, Gretherian coherency requires commutativity between distortion and Bayesian updating. Grether's $\alpha$--$\beta$ rule satisfies Gretherian coherency if $\alpha=\beta>0$, whereas it is divisible if and only if $\alpha=1$. Thus, our main result implies that if Grether's $\alpha$--$\beta$ rule satisfies both Gretherian coherency and the sender's timing indifference between static and dynamic persuasion, then the rule must be Bayesian.

The paper also relates to misspecified learning models, such as \citet{esponda2016berk}. As noted above, Bayesian updating under misspecified models can be equivalent to some non-Bayesian updating rules in terms of resulting beliefs, and vice versa \citep{bohren2023behavioral}. Concerns about model misspecification can also rationalize some systematic distortions: \cite{strzalecki2024variational} shows that posterior beliefs generated by Grether's $\alpha$--$\beta$ rule can be interpreted as those of an agent with preferences over uncertainty under posterior beliefs and concerns about model misspecification.

Several papers study the interaction between non-Bayesian updating and persuasion. The static persuasion framework with a systematically distorted receiver was introduced by \cite{de2022non}. Our main result shows that, under regularity conditions, this framework is without loss in dynamic environments if and only if the updating rule is divisible in the sense of \cite{cripps2018divisible}. Another paper in this strand is \cite{levy2022persuasion}, in which a sender uses multiple signals to persuade a receiver with correlation neglect. Correlation neglect is not a systematic distortion and is therefore outside the scope of both \cite{de2022non} and our paper.

The paper most closely related to ours is the concurrent work of \cite{azrieli2025sequential}. They focus on linear underreaction and provide examples of environments in which the sender is better off under dynamic persuasion. By contrast, our paper characterizes the updating rules under which static and dynamic persuasion yield the same value for the sender in every persuasion environment. These rules are characterized by divisibility. Another closely related paper is \cite{yang2026stochastic}. Subsequent to an earlier draft of this paper, \cite{yang2026stochastic} obtained a similar characterization under different assumptions, using a novel technique for general optimization problems in which one chooses a probability measure subject to domination by a given measure under an integral stochastic order.

\section{Framework}\label{sec:model}
  This section introduces the framework for static and dynamic persuasion with a non-Bayesian receiver.

\subsection{Environment}

Let $\Theta$ be a finite set of payoff-relevant states, and let $p \in \Delta^\circ(\Theta)$ denote a full-support prior over $\Theta$.\footnote{For every subset $X$ of a metrizable space, we denote by $\Delta(X)$ the set of Borel probability measures on $X$, endowed with the topology of weak convergence. When $X$ is finite, we identify each element of $\Delta(X)$ with a probability mass function, and let $\Delta^\circ(X)= \{q\in \Delta(X): q(x) >0 \text{ for all }x \in X\}$.} We assume $|\Theta|\geq 2$ to rule out trivial cases. Throughout the paper, when $\Theta=\{0,1\}$, we identify each $r\in\Delta(\Theta)$ with its mass on state $1$, namely $r(1)\in[0,1]$. The receiver chooses an action $a$ from a compact metrizable action set $A$.

Both the sender and the receiver are expected-utility maximizers. Their utility functions are $v:A\times\Theta\to\mathbb{R}$ and $u:A\times\Theta\to\mathbb{R}$, respectively, and both are continuous. Let $\hat{a}(q)$ denote a selection from the receiver's best-response set at belief $q\in \Delta(\Theta)$:
\[
  \hat{a}(q) \in \operatorname{BR}(q):=\arg\max_{a \in A} \mathbb{E}_{\tilde{\theta}\sim q} \left[u(a, \tilde{\theta})\right].
\]
As is standard in the persuasion literature, we assume that the receiver breaks ties in favor of the sender. Define $\hat{v}:\Delta(\Theta)\times \Delta(\Theta)\to \mathbb{R}$ by
\[
  \hat{v}(q,q') := \mathbb{E}_{\tilde{\theta}\sim q} \left[ v(\hat{a}(q'), \tilde{\theta}) \right].
\]
This is the sender's indirect utility when the sender's belief is $q$ and the receiver's belief is $q'$, so that the receiver chooses action $\hat{a}(q')$. We refer to any such tuple $(\Theta,p,A,u,v)$ as a \textbf{persuasion environment}.

  \subsection{Static persuasion}

We briefly review the static persuasion framework with a non-Bayesian receiver developed by \cite{de2022non}. The sender chooses an experiment $\sigma = \langle S, (\sigma_\theta)_{\theta \in \Theta} \rangle$, where $S$ is a finite set of signal realizations and $\sigma_\theta \in \Delta(S)$ for each $\theta \in \Theta$. Throughout the paper, we fix a sufficiently large finite signal space $S$ available to the sender. Accordingly, the set of feasible experiments is $\Sigma:=\Delta(S)^\Theta$.

The receiver observes the experiment $\sigma$ and the realized signal $s \in S$, and then chooses an action to maximize her expected utility.

We assume that the sender is Bayesian, whereas the receiver need not be. An updating rule is a function $\mu: \Sigma \times \Delta(\Theta) \to \Delta(\Theta)^S$, where $\mu(\sigma, p)(\cdot \mid s)$ denotes the posterior belief induced by signal $s \in S$, given experiment $\sigma \in \Sigma$ and prior $p \in \Delta(\Theta)$ under updating rule $\mu$. The Bayesian updating rule, denoted by $\mu_B$, is given by
\[
  \mu_B(\sigma, p)(\theta \mid s) := \frac{\sigma_\theta(s) \, p(\theta)}{\sum_{\theta' \in \Theta} \sigma_{\theta'}(s) \, p(\theta')}
\]
for all $s\in S$ such that $\sum_{\theta' \in \Theta}\sigma_{\theta'}(s)p(\theta')>0$. We assume that the sender knows the receiver's updating rule $\mu$.

Given the receiver's updating rule $\mu$ and the common prior $p$, the sender's one-shot persuasion problem is
\[
  \sup_{\tau \in T^1(p, \mu)} \mathbb{E}_{(\tilde{q},\tilde{q}')\sim\tau}[\hat{v}(\tilde{q},\tilde{q}')],
\]
where $T^1(p,\mu)$ denotes the set of feasible distributions over pairs of sender and receiver posterior beliefs induced by some experiment under prior $p$ and updating rule $\mu$.\footnote{The formal definition of $T^1(p,\mu)$ is relegated to Appendix \ref{subsec:prelim_lemmas}.}

\cite{de2022non} characterize the solution to the sender's problem for a restricted class of updating rules, called systematic distortions. Throughout the rest of the paper, for any belief $r\in\Delta(\Theta)$, let $\Theta_r:=\{\theta\in\Theta:r(\theta)>0\}$ denote the support of $r$.\footnote{For any partial-support belief $r$, we identify each element $x\in \Delta(\Theta_r)$ with the element of $\Delta(\Theta)$ satisfying $x(\theta) = 0$ for all $\theta \notin \Theta_r$.}

\begin{dfn}\label{dfn:systematic_distortion}
  An updating rule $\mu$ is a \textbf{systematic distortion} if there exists a \textbf{distortion rule} $D:= (D_p)_{p\in \Delta(\Theta)}$ such that $D_p: \Delta(\Theta_p) \to \Delta(\Theta)$ and, for any experiment $\sigma$ and any signal realization $s$ satisfying $\sum_{\theta \in \Theta}\sigma_{\theta}(s)p(\theta)>0$,
\[
  \mu(\sigma, p)(\cdot \mid s) = D_p\left( \mu_B(\sigma, p)(\cdot \mid s) \right).
\]
\end{dfn}

Under a systematic distortion, the agent forms her posterior as if she first computes the Bayesian posterior from the signal and then applies a prior-dependent distortion map.

Below, we provide several examples of updating rules that are systematic distortions.\footnote{For additional examples, see \cite{de2022non}. The non-Bayesian updating rules listed in Example 1 of \cite{bohren2023behavioral} are also systematic distortions.}

\begin{ex}\label{ex:grether}
  \textbf{Grether's $\alpha$--$\beta$ rule} \citep{grether1980bayes}: For $\alpha > 0$ and $\beta > 0$,
  \[
    \mu(\sigma, p)(\theta \mid s) = \frac{p(\theta)^\alpha \, \sigma_\theta(s)^\beta}{\sum_{\theta' \in \Theta} p(\theta')^\alpha \, \sigma_{\theta'}(s)^\beta}
  \]
  for each $\theta \in \Theta$.
  The parameter $\alpha$ captures distortion in the prior, or base rates, while $\beta$ captures distortion in the signal distribution conditional on the state. If $\alpha \in (0,1)$, the agent underweights prior information; we refer to this as \textbf{base-rate neglect}. If $\alpha > 1$, the agent overweights prior information; we refer to this as \textbf{base-rate overweighting}. If $\beta \in (0,1)$, the agent underweights new information; we refer to this as \textbf{geometric underreaction}. If $\beta > 1$, we refer to this as \textbf{geometric overreaction}. When $\alpha=1$, we refer to Grether's rule as \textbf{geometric distortion}; Bayesian updating is the special case $\beta=1$.

  The associated distortion rule is
  \[
    D_p(q)(\theta) = \begin{cases}
    \frac{p(\theta)^{\alpha - \beta} q(\theta)^\beta}{\sum_{\theta' \in \Theta_p} p(\theta')^{\alpha - \beta} q(\theta')^\beta} & \text{if } \theta \in \Theta_p,\\
    0 & \text{if } \theta \notin \Theta_p.
    \end{cases}
  \]
\end{ex}
\begin{ex}\label{ex:linear_underreaction}
\textbf{Linear underreaction:} For some $\lambda \in [0,1]$,
  \[
    \mu(\sigma,p)(\theta\mid s) = \lambda\frac{ p(\theta) \sigma_{\theta}(s)}{\sum_{\theta'\in \Theta}p(\theta')\sigma_{\theta'}(s) } + (1 - \lambda)p(\theta)
  \]
  for each $\theta\in \Theta$. If $\lambda = 1$, the rule reduces to Bayesian updating. If $\lambda \in [0,1)$, the agent discounts the Bayesian posterior and mixes it with the prior; we refer to this distortion as \textbf{linear underreaction}. The associated distortion rule is
  \[
    D_p(q)(\theta) = \lambda q(\theta) + (1-\lambda)p(\theta).
  \]
\end{ex}

Throughout the remainder of the paper, we focus on systematic distortions. When the receiver's updating rule $\mu$ admits distortion rule $D$, we often write $T^1(p,D)$ instead of $T^1(p,\mu)$.

Building on the approach of \cite{alonso2016bayesian} for Bayesian persuasion with heterogeneous priors, \cite{de2022non} show that, if the receiver's distortion rule is $D$, then for any full-support prior $p\in \Delta^\circ(\Theta)$, the sender's problem becomes
\[
  V(p, D) = \sup_{\rho \in \mathcal{R}(p)} \mathbb{E}_{\tilde{q}\sim\rho}[\check{v}^{D_p}(\tilde{q})] = \left[\operatorname{CAV}(\check{v}^{D_p})\right](p),
\]
where $\mathcal{R}(p) := \left\{ \rho \in \Delta(\Delta(\Theta)) : \mathbb{E}_{\tilde{q}\sim\rho}[\tilde{q}] = p \right\}$ is the set of Bayes-plausible distributions over posteriors given prior $p$, $\check{v}^{D_p}(q) := \hat{v}(q,D_p(q))$ is the sender's distorted indirect utility, and
\[
  \left[\operatorname{CAV}(f)\right](p) := \sup\{ z \mid (p, z) \in \operatorname{conv}(\operatorname{Gr}(f)) \}
\]
denotes the concavification of $f$ at $p$, where $\operatorname{conv}(\operatorname{Gr}(f))$ is the convex hull of the graph of $f$. Thus, the problem can be recast as a standard Bayesian persuasion problem in the sense of \cite{kamenica2011bayesian}, but with a modified indirect utility function.

\subsection{Dynamic persuasion}

We now extend the framework to dynamic persuasion. We focus on a parsimonious two-step model.\footnote{We discuss the extension to $n\geq 3$ steps in Section \ref{sec:discussion}.} The process unfolds as follows:\footnote{Since the receiver acts only at the end, it does not matter whether the sender commits at the outset to both $\sigma^1$ and $\{\sigma^2_{s_1}\}_{s_1\in S}$, or instead commits to $\sigma^1$ at the beginning of the first stage and then chooses $\{\sigma^2_{s_1}\}_{s_1\in S}$ contingent on the first-stage signal realization $s_1$.}
\begin{enumerate}
  \item The sender chooses a first-step experiment $\sigma^1\in \Sigma$ and a collection of second-step experiments $\{\sigma^2_{s_1}\}_{s_1\in S}\in\Sigma^S$, where $\sigma^2_{s_1}$ is used after the first-stage signal realization $s_1$.
  \item After observing $\sigma^1$ and its realization $s_1$, the receiver updates her belief from $p$ to $D_p(q)$, where $q:=\mu_B(\sigma^1,p)(\cdot\mid s_1)$.
  \item After observing $\sigma^2_{s_1}$ and its realization $s_2$, the receiver updates her belief from $D_p(q)$ to
  \[
    D_{D_p(q)}\left(\mu_B\left(\sigma^2_{s_1},D_p(q)\right)(\cdot\mid s_2)\right).
  \]
  \item The receiver chooses an action $a \in A$.
\end{enumerate}
  The sender's dynamic persuasion problem is
\[
  \sup_{\tau\in T^2(p,D)} \mathbb{E}_{(\tilde{r}, \tilde{r}')\sim\tau}[\hat{v}(\tilde{r}, \tilde{r}')],
\]
where $T^2(p,D)$ denotes the set of feasible joint distributions over the sender's and receiver's final posterior beliefs induced by some dynamic persuasion strategy $(\sigma^1,\{\sigma^2_{s_1}\}_{s_1\in S})$ under prior $p$ and distortion rule $D$.\footnote{The formal definition of $T^2(p,D)$ is relegated to Appendix \ref{subsec:prelim_lemmas}.}

Let
\[
  \mathcal{V}^1(p,D) := \left\{\mathbb{E}_{(\tilde{r}, \tilde{r}')\sim\tau}[\hat{v}(\tilde{r}, \tilde{r}')]: \tau \in T^1(p,D)\right\}
\]
and
\[
  \mathcal{V}^2(p,D) := \left\{\mathbb{E}_{(\tilde{r}, \tilde{r}')\sim\tau}[\hat{v}(\tilde{r}, \tilde{r}')]: \tau \in T^2(p,D)\right\}
\]
be the sets of feasible ex ante sender payoffs under static and dynamic persuasion, respectively.



\section{Motivating Example: Dynamic Persuasion with Grether's $\alpha$--$\beta$ Rule}\label{sec:non_divisible_multi_step}

In this section, we show that, with a non-Bayesian receiver, the sender's value may differ under static and dynamic persuasion.

To illustrate, consider the persuasion environment $(\Theta, p, A, u, v)$ with $\Theta = \{0,1\}$, $p\in (0,1/2)$, $A= \{0,1\}$, $u(a,\theta) = \mathbbm{1}_{a = \theta}$, and $v(a,\theta) = a$ for all $\theta$. We call this the \textbf{judge-prosecutor environment}, as it corresponds to the classic judge-prosecutor example in \cite{kamenica2011bayesian}.

  When the receiver follows Grether's $\alpha$--$\beta$ rule in this environment, the following result obtains.

\begin{prop}\label{prop:grether}
  Consider the judge-prosecutor environment with a receiver who updates according to Grether's $\alpha$--$\beta$ rule. Then, for any $\beta > 0$,
  \begin{enumerate}
    \item If $\alpha\in (0,1)$, the sender obtains a higher value under dynamic persuasion:
    \[
      \sup_{x\in \mathcal{V}^2(p, D)} x > \sup_{x\in \mathcal{V}^1(p, D)} x.
    \]
    \item If $\alpha \geq 1$, the sender is indifferent between static and dynamic persuasion:
    \[
      \sup_{x\in \mathcal{V}^2(p, D)} x = \sup_{x\in \mathcal{V}^1(p, D)} x.
    \]
  \end{enumerate}
\end{prop}

The intuition is as follows. Recall that the receiver's best response is $\hat{a}(r') = \mathbbm{1}_{r' \geq \frac{1}{2}}$. Thus, $1/2$ is the lowest receiver posterior at which the sender's preferred action, $a=1$, is optimal. Under base-rate neglect, $\alpha < 1$, even if the sender provides no information in the first step, the receiver's belief drifts toward the uniform prior before the second step, which is more favorable to the sender. Because this distorted prior is closer to the threshold $1/2$, the sender can obtain a higher payoff in the second step.
  
  Under base-rate overweighting, $\alpha > 1$, by contrast, if the sender reveals no information in the first step, then before the second step the receiver's belief drifts farther from the threshold: $D_p(p)<p<1/2$. This makes persuasion more difficult. In this case, however, the optimal static payoff is still attainable under dynamic persuasion. If the sender induces an intermediate receiver belief satisfying either $D_p(q)=0$ or $D_p(q)=1/2$ and then reveals no information in the second step, the receiver's belief remains unchanged, so the optimal one-shot payoff is attained.

  Notably, Proposition \ref{prop:grether} is independent of $\beta>0$. This is natural in light of Corollary~\ref{cor:geometric_distortion}: geometric over- or underweighting of signal likelihoods does not affect the value of delaying information delivery. Geometric overreaction and underreaction are divisible updating rules, which we characterize as rules under which the timing of information delivery does not affect the sender's value.

  To prove Proposition \ref{prop:grether}, we use a technique similar to that in \cite{alonso2016bayesian} to solve the sender's \textit{interim problem}: choosing the second-step experiment $\sigma^2_{s_1}\in\Sigma$ contingent on the first-step experiment $\sigma^1$ and realization $s_1$. This is a heterogeneous non-Bayesian persuasion problem because, when $\sigma^1$ and $s_1$ induce sender belief $q$, the receiver's belief is $D_p(q)$, and these beliefs serve as the relevant ``priors'' in the interim problem. The interim problem can be recast as one with common prior $q$ and distortion rule $D^{II}(\cdot;p)$ defined by
\begin{equation}\label{eqn:2nd_distortion}
  D_{q}^{II}(r;p) 
  := D_{D_p(q)}\left(\mathcal{B}\left(\frac{D_p(q)}{q},r\right) \right),
\end{equation}
for any full-support beliefs $p,q,r$ such that $D_p(q)$ also has full support,\footnote{With Grether's $\alpha$--$\beta$ rule, this transformation extends to partial-support beliefs as well.} where
\[
  \mathcal{B}\left(\frac{D_p(q)}{q},r\right) = \frac{\frac{D_p(q)}{q}r}{\left\langle \frac{D_p(q)}{q}, r \right\rangle}.\footnote{Unless otherwise noted, for any $x,y\in\mathbb{R}^{|\Theta|}$, we write $xy:=\left(x(\theta)y(\theta)\right)_{\theta\in \Theta}$ for the coordinatewise product. For any $x\in\mathbb{R}^{|\Theta|}$ and $y\in \mathbb{R}_{++}^{|\Theta|}$, we write $\frac{x}{y}:=\left(\frac{x(\theta)}{y(\theta)}\right)_{\theta\in \Theta}$ for the coordinatewise ratio. Finally, for any $x,y\in\mathbb{R}^{|\Theta|}$, let $\langle x,y\rangle := \sum_{\theta\in \Theta} x(\theta)y(\theta)$.}
\]

For a formal statement of this transformation lemma, see Lemma \ref{lem:hetero_transformation} in the Appendix.


  Therefore, for any given $q$, the interim problem can be viewed as a static persuasion problem with distortion rule $D^{II}(\cdot;p)$ in the sense of \cite{de2022non}. Hence, the sender's interim value is given by the concavification of the distorted indirect utility $\check{v}^{D_q^{II}(\cdot;p)}$ evaluated at $q$. The sender's value in the overall dynamic problem is obtained by concavifying this interim value once more. The proof of Proposition \ref{prop:grether} is relegated to Section \ref{app:proof_grether} of the appendix.

  \section{Characterization Results}\label{sec:characterization}

While the preceding approach allows us to solve the dynamic persuasion problem, it may become analytically intractable because it replaces the original distortion rule with an interim one. For example, under linear underreaction, the modified distortion rule $D^{II}(\cdot;p)$ is analytically cumbersome.

  Given this difficulty, an analyst may prefer the more tractable one-shot persuasion framework of \cite{de2022non} to the dynamic framework. When is this without loss? To answer this question, we impose the following mild regularity conditions on the receiver's distortion rule.
  \begin{dfn}\label{dfn:regular_D}
  A distortion rule $D$ is \textbf{regular} if it satisfies:
  \begin{enumerate}
    \item \textbf{Non-decreasing support}: $\Theta_q \subseteq \Theta_{D_p(q)}$ for all $p \in \Delta^\circ(\Theta)$ and $q \in \Delta(\Theta)$.
    \item \textbf{Continuity}: The map $(p,q)\mapsto D_p(q)$ from $\Delta^\circ(\Theta)\times\Delta(\Theta)$ to $\Delta(\Theta)$ is continuous.
    \item \textbf{Injectivity}: $D_p$ is one-to-one for every $p \in \Delta^\circ(\Theta)$.
  \end{enumerate}
\end{dfn}

 Grether's $\alpha$--$\beta$ rule and linear underreaction in Examples \ref{ex:grether} and \ref{ex:linear_underreaction} satisfy these regularity conditions. The conditions are imposed for technical reasons but are also natural.

In the next theorem, we also impose the following axiom.

\begin{dfn}\label{dfn:prior_independence_under_certainty}
  A distortion rule $D$ satisfies \textbf{prior independence under certainty} if, for any full-support beliefs $p,q$,
  \[
    D_p(\delta_\theta) = D_{D_p(q)}(\delta_\theta) \quad \text{for all } \theta \in \Theta_{D_p(q)}.
  \]
\end{dfn}

Roughly speaking, this axiom requires that, after a fully revealing signal, the agent's belief is uniquely determined regardless of the prior. The axiom is natural and is satisfied by Grether's $\alpha$--$\beta$ rule. However, linear underreaction does not satisfy it for any $\lambda \in [0,1)$. Later in this section, we impose a different axiom, which is satisfied by linear underreaction, and obtain a similar characterization. 

We now introduce divisibility, the key condition for the sender's indifference between static and dynamic persuasion. The modified distortion rule $D^{II}(\cdot;p)$ defined in \eqref{eqn:2nd_distortion} again plays a central role.
\begin{dfn}\label{dfn:divisibility}
  A regular distortion rule $D$ is \textbf{divisible} if, for any full-support beliefs $p$, $q$, and $r$, 
  \[
    D_p(r)=D_q^{II}(r;p).
  \]
\end{dfn}
Under a divisible updating rule, given two signals, the posterior from simultaneous updating (the left-hand side) coincides with that from sequential updating (the right-hand side). This is the analogue of the divisibility axiom in \cite{cripps2018divisible}, expressed in terms of distortion rules.

We now characterize divisible distortion rules in terms of the sender's indifference between static and dynamic persuasion:

\begin{thm}\label{thm:divisible_iff_indifference}
  Suppose the receiver's distortion rule $D$ is regular and satisfies prior independence under certainty. Then the following are equivalent:
  \begin{enumerate}
    \item The distortion rule $D$ is divisible.
    \item The sets of feasible ex ante sender payoffs coincide under static and dynamic persuasion in every persuasion environment:
    \[
      \mathcal{V}^1(p,D)=\mathcal{V}^2(p,D)\quad \text{for every } (\Theta,p,A,u,v).
    \]
    \item The sender is indifferent between static and dynamic persuasion in every persuasion environment:
    \[
      \sup_{x\in\mathcal{V}^1(p,D)} x=\sup_{x\in\mathcal{V}^2(p,D)} x
      \quad \text{for every } (\Theta,p,A,u,v).
    \]
  \end{enumerate}
\end{thm}
\textit{Proof Sketch}. To see the argument, consider the following lemma.

\begin{lem}\label{lem:T1_equal_T2_iff_divisible}
  Suppose the receiver's distortion rule $D$ is regular. Then $D$ is divisible if and only if $T^1(p,D) = T^2(p,D)$ for every $\Theta$ and every full-support prior $p$ over $\Theta$.
\end{lem}
Lemma \ref{lem:T1_equal_T2_iff_divisible} characterizes divisibility by the coincidence of the sets of feasible joint distributions over the sender's and receiver's final posterior beliefs under static and dynamic persuasion. Given this lemma, \(1\implies 2\) follows immediately, and \(2\implies 3\) is immediate. The nontrivial part is \(3\implies 1\). If the sender's indirect utility could be any bounded continuous function of the joint posterior pair, equality of values in all environments would imply \(T^1(p,D)=T^2(p,D)\) by a separating hyperplane argument, where the separating functional is
\[
  \tau \mapsto \int_{\Delta(\Theta)\times\Delta(\Theta)} \hat{v}\,d\tau .
\]
In our model, however, \(\hat v(q,q')\) is affine in the sender's posterior \(q\), since the sender is an expected-utility maximizer, so this direct argument fails. Instead, we use environments in which the receiver truthfully reports her posterior under a proper scoring rule and the sender's payoff is state-independent. In such environments, the sender's indirect utility can be any bounded continuous function of the receiver's posterior, that is, any bounded continuous function from $\Delta(\Theta)$ to $\mathbb{R}$. A separating hyperplane argument with this reduced indirect utility yields divisibility up to an affine transformation. Prior independence under certainty and injectivity of \(D_p\) then force this transformation to be the identity, yielding divisibility.

\vspace{\baselineskip}
Theorem \ref{thm:divisible_iff_indifference} shows that, for any divisible distortion rule, the tractable static persuasion framework of \cite{de2022non} is without loss relative to dynamic persuasion.

With an additional approachability condition, \cite{cripps2018divisible} provides a functional-form characterization of divisible distortion rules. In particular, if for some full-support prior $p^*$ the restriction
\[
  D_{p^*}\big|_{\Delta^\circ(\Theta)}:\Delta^\circ(\Theta)\to\Delta^\circ(\Theta)
\]
is onto, then the distortion rule is divisible if and only if, for some homeomorphism $F:\Delta^\circ(\Theta)\to\Delta^\circ(\Theta)$, for all full-support beliefs $p$ and $q$,
\[
  D_p(q)
  =
  F^{-1}\left(
    \frac{\frac{F(p)}{p}q}{\left\langle \frac{F(p)}{p}, q \right\rangle}
  \right).\footnote{The formal statement is in Section \ref{app:divisible_functional_form} of the Appendix.}
\]
 This class includes Grether's $\alpha$--$\beta$ rule with $\alpha = 1$, that is, geometric distortion. Therefore, we have the following corollary.

\begin{cor}\label{cor:geometric_distortion}
  If the receiver's distortion rule is a geometric distortion, then the sender is indifferent between static and dynamic persuasion in every persuasion environment.
\end{cor}

Grether's $\alpha$--$\beta$ rule has been widely used in the literature on non-Bayesian updating because of its tractability. Corollary \ref{cor:geometric_distortion} shows that, within Grether's $\alpha$--$\beta$ class, the restriction to geometric distortion allows dynamic information arrival to be analyzed using the tractable static framework without loss.

\subsection{Characterization without Prior Independence under Certainty}

Theorem \ref{thm:divisible_iff_indifference} characterizes divisibility by the sender's timing indifference in every persuasion environment, but it relies on prior independence under certainty, which some commonly used distortion rules, such as linear underreaction, do not satisfy. Below, we impose a different axiom and characterize updating rules under which the timing of information delivery does not affect the sender's value.

\begin{dfn}\label{dfn:invariance_under_no_information}
  A distortion rule $D$ satisfies \textbf{invariance under no information} if, for every full-support belief $p$, $D_p(p) = p$.
\end{dfn}
The interpretation is simple: when the agent receives an uninformative signal, so that the Bayesian posterior equals the prior, the agent's post-distortion belief remains equal to that prior. This axiom is satisfied by linear underreaction and geometric distortion.

With this axiom, we prove a characterization theorem similar to Theorem \ref{thm:divisible_iff_indifference} by focusing on a subclass of persuasion environments. If the sender's payoff depends only on the receiver's action, that is, if $v(a,\theta)$ is constant in $\theta$ for each $a\in A$, we say that the environment has \textbf{state-independent sender preferences}. Below, we characterize the sender's indifference between static and dynamic persuasion in every such environment.

The desired characterization uses a weakened version of divisibility, which we call \textit{weak divisibility}.

\begin{dfn}\label{dfn:weak_divisibility}
  A regular distortion rule $D$ is \textbf{weakly divisible} if it satisfies invariance under no information and, for all full-support beliefs $p,q$, there exists a full-rank column-stochastic matrix $M_{p,q}$ such that
  \[
    D_p(M_{p,q} r) = D^{II}_q(r;p)
    \quad \text{for every } r\in\Delta(\Theta).
  \]
\end{dfn}

Compared with divisibility, weak divisibility requires invariance under no information but allows inconsistency between the first- and second-step distortions from the sender's perspective up to an affine transformation. Under this relaxed condition, we still obtain the sender's indifference between static and dynamic persuasion, as the following theorem shows.

\begin{thm}\label{thm:weakly_divisibile_iff_indifference}
  Suppose the receiver has a regular distortion rule $D$ satisfying invariance under no information. Then the following are equivalent:
  \begin{enumerate}
    \item The distortion rule $D$ is weakly divisible.
    \item The feasible sets of ex ante sender payoffs coincide under static and dynamic persuasion in every persuasion environment with state-independent sender preferences.
    \item The sender is indifferent between static and dynamic persuasion in every persuasion environment with state-independent sender preferences.
  \end{enumerate}
\end{thm}



Linear underreaction is not weakly divisible. Since invariance under no information implies that the sender's value under dynamic persuasion is weakly higher than under static persuasion, Theorem \ref{thm:weakly_divisibile_iff_indifference} implies that, in some persuasion environment, the sender strictly prefers dynamic persuasion.\footnote{The formal statement and proof are in Section \ref{app:linear_underreaction} of the Appendix.}

  \section{Discussion}\label{sec:discussion}

\subsection{Multiple-Step Persuasion}

We have focused on a simple two-step setting to study dynamic non-Bayesian persuasion. The model can be extended to $n$-step persuasion for $n\geq 3$ by extending Lemma \ref{lem:hetero_transformation} accordingly. As long as the receiver's distortion rule is divisible, the static persuasion approach remains without loss.

One can also study $n$-step persuasion with non-divisible distortion rules. If the distortion rule satisfies invariance under no information (Definition \ref{dfn:invariance_under_no_information}), then the sender's value is weakly increasing in $n$, because the sender can always choose an uninformative experiment in any additional step. However, characterizing an optimal persuasion strategy is often analytically intractable.

Beyond this analytical complexity, some commonly observed rules may imply extreme outcomes that seem unnatural in certain contexts. For example, consider the judge-prosecutor environment in Section \ref{sec:non_divisible_multi_step} under Grether's $\alpha$--$\beta$ rule with $\alpha<1$. Suppose the sender uses uninformative experiments through step $n-1$ and then uses an optimal experiment at step $n$. As $n\to\infty$, the receiver's prior at the beginning of step $n$ converges to the uniform distribution, which is exactly the threshold belief for the sender's preferred action. Thus, for sufficiently large $n$, the sender can obtain a payoff arbitrarily close to $1$, regardless of the initial prior $p\in(0,1/2)$.\footnote{Grether's $\alpha$--$\beta$ rule with $\alpha\neq 1$ violates invariance under no information. Hence, the monotonicity argument based on adding uninformative experiments does not apply to this example in general environments.} Avoiding such extreme outcomes may require restricting the horizon $n$ or considering alternative distortion rules.

  \subsection{Non-Expected Utilities}\label{sec:belief_dependent_sender_payoff}

As noted in the proof sketch of Theorem \ref{thm:divisible_iff_indifference}, the fact that the sender's indirect utility $\hat{v}$ is affine in the sender's belief precludes a straightforward use of the separating hyperplane argument to prove \(3\implies 1\). If the persuasion environment is extended to allow any belief-dependent sender utilities, then, by varying the environment, the sender's indirect utility can potentially be any bounded continuous function. As a result, the separating hyperplane argument can then be applied directly. Thus, in this extended model, regularity conditions alone are needed to characterize the sender's indifference between static and dynamic persuasion by divisibility of the distortion rule.\footnote{Subsequent to an earlier version of this paper, \cite{yang2026stochastic} provide such a characterization.}

  \appendix

\section{Proofs and Lemmas}\label{sec:proofs}

\subsection{Preliminary Results}\label{subsec:prelim_lemmas}

We use the following lemmas to prove the paper's main results.

\begin{lem}\label{lem:hetero_transformation}
  Suppose the receiver's distortion rule $D$ satisfies the non-decreasing support property in Definition \ref{dfn:regular_D}. Fix any persuasion environment $(\Theta,p,A,u,v)$, any dynamic persuasion strategy
  $\left(\sigma^1,\{\sigma^2_{s_1}\}_{s_1\in S}\right)\in \Sigma\times\Sigma^S$, and any signal pair $(s_1,s_2)\in S^2$
  such that $\sum_{\theta\in \Theta} p(\theta)\sigma^1_\theta(s_1)\sigma^2_{s_1,\theta}(s_2)>0$. Suppose that either
  \begin{enumerate}
    \item $q:=\mu_B(\sigma^1,p)(\cdot\mid s_1)$ has full support, or
    \item $q$ and $D_p(q)$ have the same support, that is, $\Theta_q=\Theta_{D_p(q)}$.
  \end{enumerate}
  Then the receiver's final belief in dynamic persuasion is
  \begin{equation}\label{eqn:final_receiver_belief_lem_hetero_transformation}
    D_q^{II}(r;p)=D_{D_p(q)}\left(\mathcal{B}\left(\frac{D_p(q)}{q},r\right)\right),
  \end{equation}
  where $r = \mu_B(\sigma^2_{s_1},q)(\cdot\mid s_2)$ and
  \[
    \mathcal{B}\left(\frac{D_p(q)}{q},r\right)
    =
    \frac{\frac{D_p(q)}{q}r}{\left\langle \frac{D_p(q)}{q}, r \right\rangle}.\footnote{When $\frac{D_p(q)(\theta)}{q(\theta)} = \frac{0}{0}$ for some $\theta\in \Theta$, we adopt the convention that $0/0 = 0$.}
  \]
  
\end{lem}
\begin{proof}
  Fix $D$, $(\Theta,p,A,u,v)$, a dynamic persuasion strategy
$\left(\sigma^1,\{\sigma^2_{s_1}\}_{s_1\in S}\right)\in\Sigma\times\Sigma^S$, and a signal pair $(s_1,s_2)\in S^2$ satisfying the conditions in the lemma. Since the receiver's updating rule is a systematic distortion with distortion rule $D$, the receiver's final belief is
\begin{equation}\label{eqn:proof_hetero_2}
  D_{D_p(q)}
  \bigl(\mu_B(\sigma^2_{s_1},D_p(q))(\cdot\mid s_2)\bigr)
  =
  D_{D_p(q)}
  \left(
  \left(
  \frac{D_p(q)(\theta^*)\sigma^2_{s_1,\theta^*}(s_2)}
  {\sum_{\theta'\in \Theta} D_p(q)(\theta')\sigma^2_{s_1,\theta'}(s_2)}
  \right)_{\theta^*\in \Theta}
  \right),
\end{equation}
where the equality follows from Bayes' rule.

Define
\[
  r:=\mu_B(\sigma^2_{s_1},q)(\cdot\mid s_2).
\]
By Bayes' rule, for every $\theta\in\Theta$,
\[
  r(\theta)
  =
  \frac{q(\theta)\sigma^2_{s_1,\theta}(s_2)}
  {\sum_{\theta'\in \Theta}q(\theta')\sigma^2_{s_1,\theta'}(s_2)}.
\]
Equivalently,
\[
  q(\theta)\sigma^2_{s_1,\theta}(s_2)
  =
  r(\theta)
  \sum_{\theta'\in \Theta}q(\theta')\sigma^2_{s_1,\theta'}(s_2).
\]

First suppose that $q$ has full support. Multiplying both sides by $D_p(q)(\theta)/q(\theta)$ yields
\begin{equation}\label{eqn:proof_hetero_1}
  D_p(q)(\theta)\sigma^2_{s_1,\theta}(s_2)
  =
  \frac{D_p(q)(\theta)}{q(\theta)}r(\theta)
  \sum_{\theta'\in \Theta}q(\theta')\sigma^2_{s_1,\theta'}(s_2).
\end{equation}
Substituting \eqref{eqn:proof_hetero_1} into the coordinate formula for
$\mu_B(\sigma^2_{s_1},D_p(q))(\cdot\mid s_2)$ and cancelling the common positive factor
$\sum_{\theta'\in \Theta}q(\theta')\sigma^2_{s_1,\theta'}(s_2)$ from numerator and denominator gives
\[
  \mu_B(\sigma^2_{s_1},D_p(q))(\cdot\mid s_2)
  =
  \mathcal{B}\left(\frac{D_p(q)}{q},r\right).
\]
Therefore,
\[
  D_{D_p(q)}
  \bigl(\mu_B(\sigma^2_{s_1},D_p(q))(\cdot\mid s_2)\bigr)
  =
  D_q^{II}(r;p).
\]
By non-decreasing support, if $q$ has full support, then so does $D_p(q)$, so $D_q^{II}(r;p)$ is well-defined.

Now suppose instead that $\Theta_q=\Theta_{D_p(q)}$. For every $\theta\in\Theta_q$, equation \eqref{eqn:proof_hetero_1} remains valid. For every $\theta\notin\Theta_q$, we have $D_p(q)(\theta)=0$ and $r(\theta)=0$. Under the convention $0/0=0$, equation \eqref{eqn:proof_hetero_1} therefore holds for all $\theta\in\Theta$. Substituting these identities into the coordinate formula for
$\mu_B(\sigma^2_{s_1},D_p(q))(\cdot\mid s_2)$ again yields
\[
  \mu_B(\sigma^2_{s_1},D_p(q))(\cdot\mid s_2)
  =
  \mathcal{B}\left(\frac{D_p(q)}{q},r\right).
\]
Hence,
\[
  D_{D_p(q)}
  \bigl(\mu_B(\sigma^2_{s_1},D_p(q))(\cdot\mid s_2)\bigr)
  =
  D_q^{II}(r;p),
\]
as desired.
\end{proof}

\bigskip

  Consider any persuasion environment $(\Theta,p,A,u,v)$ and any receiver distortion rule $D$. The sets of feasible joint distributions of the sender's and receiver's final posterior beliefs under static and dynamic persuasion are
\[
\begin{aligned}
T^1(p,D):=\Bigl\{\tau\in\Delta(\Delta(\Theta)\times\Delta(\Theta)) :
&\ \exists\sigma\in \Sigma \text{ such that, for all } (q,q'), \\
&\ \tau(q,q')=\sum_{\theta\in\Theta}p(\theta)
\sum_{s\in S^1(q,q';\sigma,p,D)}\sigma_\theta(s)
\Bigr\},
\end{aligned}
\]
where
\[
  S^1(q,q';\sigma,p,D)
  :=
  \{s\in S:\mu_B(\sigma,p)(\cdot\mid s)=q \text{ and } D_p(q)=q'\}.
\]
Similarly,
\[
\begin{aligned}
T^2(p,D):=\Bigl\{&\tau\in\Delta(\Delta(\Theta)\times\Delta(\Theta)) :
\ \exists\sigma^1\in\Sigma,\ \{\sigma^2_{s_1}\}_{s_1\in S}\in\Sigma^S
\text{ such that, for all } (r,r'),\\
&\ \tau(r,r')=
\sum_{\theta\in\Theta}p(\theta)
\sum_{s_1\in S}\sigma^1_\theta(s_1)
\sum_{s_2\in S^2(r,r';s_1,\sigma^1,\{\sigma^2_{s_1}\}_{s_1\in S},p,D)}
\sigma^2_{s_1,\theta}(s_2)
\Bigr\},
\end{aligned}
\]
where
\[
\begin{aligned}
&S^2(r,r';s_1,\sigma^1,\{\sigma^2_{s_1}\}_{s_1\in S},p,D) \\
&\quad :=
\Bigl\{s_2\in S:
r = \mu_B(\sigma^2_{s_1}, q)(\cdot\mid s_2),\\
&r' = D_{D_p(q)}\bigl(\mu_B(\sigma^2_{s_1},D_p(q))(\cdot\mid s_2)\bigr),\
q = \mu_B(\sigma^1,p)(\cdot\mid s_1)
\Bigr\}.
\end{aligned}
\]
Below, we use the following set of feasible joint distributions of the sender's and receiver's final posterior beliefs induced by a restricted class of dynamic persuasion strategies:
\[
\begin{aligned}
T^2_0(p,D) := \Bigl\{&\tau\in\Delta(\Delta(\Theta)\times\Delta(\Theta)) :
\ \exists\sigma^1\in\Sigma,\ \{\sigma^2_{s_1}\}_{s_1\in S}\in\Sigma^S \text{ such that} \\
&\text{for all }s_1\in S,\ \text{if } \sum_{\theta\in\Theta}p(\theta)\sigma^1_\theta(s_1) > 0,
   \text{ then } \sigma^1_\theta(s_1)>0 \text{ for all }\theta\in\Theta_p, \\
&\text{and, for all }(r,r'), \\
&\ \tau(r,r')=
\sum_{\theta\in\Theta}p(\theta)
\sum_{s_1\in S}\sigma^1_\theta(s_1)
\sum_{s_2\in S^2(r,r';s_1,\sigma^1,\{\sigma^2_{s_1}\}_{s_1\in S},p,D)}
\sigma^2_{s_1,\theta}(s_2)
\Bigr\}.
\end{aligned}
\]
The distributions in $T_0^2(p,D)$ are induced by dynamic persuasion strategies such that every first-step signal realization occurring with positive probability induces a full-support Bayesian posterior.

\bigskip

\begin{lem}\label{lem:T20_dense_in_T2}
  For any finite set $\Theta$, full-support prior $p\in \Delta^\circ(\Theta)$, and regular receiver distortion rule $D$, $T_0^2(p,D)$ is dense in $T^2(p,D)$ under the topology of weak convergence.
\end{lem}
\begin{proof}
  
Fix any $\tau\in T^2(p,D)$. We show that there exists a sequence in $T^2_0(p,D)$ that converges weakly to $\tau$.

By definition of $T^2(p,D)$, there exist $\sigma^1\in\Sigma$ and $\{\sigma^2_{s_1}\}_{s_1\in S}\in\Sigma^S$ such that $\tau$ is induced by $(\sigma^1,\{\sigma^2_{s_1}\}_{s_1\in S})$. Fix such a dynamic persuasion strategy.

For each $s_1\in S$, define
\[
  \pi(s_1):=\sum_{\theta\in\Theta}p(\theta)\sigma^1_\theta(s_1),
\]
and let $S_1^+ := \{s_1\in S:\pi(s_1)>0\}$. For each $s_1\in S_1^+$, let
\[
  q_{s_1}:=\mu_B(\sigma^1,p)(\cdot\mid s_1).
\]

Next, for each $s_1\in S_1^+$ and $s_2\in S$, define
\[
  \pi(s_1,s_2):=\sum_{\theta\in\Theta}p(\theta)\sigma^1_\theta(s_1)\sigma^2_{s_1,\theta}(s_2),
\]
and let
\[
  S_2^+ := \{(s_1,s_2)\in S\times S:\pi(s_1,s_2)>0\}.
\]
For each $(s_1,s_2)\in S_2^+$, define
\[
  r_{s_1,s_2}:=\mu_B(\sigma^2_{s_1},q_{s_1})(\cdot\mid s_2)
\]
and
\[
  r'_{s_1,s_2}:=
  D_{D_p(q_{s_1})}\!\left(\mu_B(\sigma^2_{s_1},D_p(q_{s_1}))(\cdot\mid s_2)\right).
\]
Then, for every bounded continuous function $f:\Delta(\Theta)\times\Delta(\Theta)\to\mathbb{R}$,
\[
  \int f\,d\tau
  =
  \sum_{(s_1,s_2)\in S_2^+}\pi(s_1,s_2)\,
  f(r_{s_1,s_2},r'_{s_1,s_2}).
\]


    We now construct a sequence in $T^2_0(p,D)$ that converges weakly to $\tau$. Fix any full-support distribution $m\in\Delta(S)$; that is, $m(s)>0$ for every $s\in S$. For each $\varepsilon\in(0,1)$, define a perturbed first-step experiment $\sigma^{1,\varepsilon}\in\Sigma$ by
\[
  \sigma^{1,\varepsilon}_\theta(s_1)
  :=
  (1-\varepsilon)\sigma^1_\theta(s_1)+\varepsilon m(s_1)
  \qquad
  \text{for all } \theta\in\Theta,\ s_1\in S.
\]
For each $\theta\in\Theta$, $\sum_{s_1\in S}\sigma^{1,\varepsilon}_\theta(s_1)=1$, and $\sigma^{1,\varepsilon}_\theta(s_1)\ge 0$ for every $s_1\in S$. Hence $\sigma^{1,\varepsilon}\in\Sigma$. Let $\tau^\varepsilon$ be the distribution induced by $\bigl(\sigma^{1,\varepsilon},\{\sigma^2_{s_1}\}_{s_1\in S}\bigr)$.

We claim that $\tau^\varepsilon\in T^2_0(p,D)$ for every $\varepsilon\in(0,1)$. Indeed, for any $s_1\in S$ such that
\[
  \pi^\varepsilon(s_1):=\sum_{\theta\in\Theta}p(\theta)\sigma^{1,\varepsilon}_\theta(s_1)>0,
\]
and any $\theta\in\Theta$, we have
\[
  \sigma^{1,\varepsilon}_\theta(s_1)
  =
  (1-\varepsilon)\sigma^1_\theta(s_1)+\varepsilon m(s_1)
  \ge \varepsilon m(s_1)>0.
\]
Hence the defining condition of $T^2_0(p,D)$ is satisfied.

For later use, note that
\[
  \pi^\varepsilon(s_1)
  =
  \sum_{\theta\in\Theta}p(\theta)\sigma^{1,\varepsilon}_\theta(s_1)
  =
  (1-\varepsilon)\pi(s_1)+\varepsilon m(s_1),
\]
so $\pi^\varepsilon(s_1)>0$ for every $s_1\in S$.

Thus, for every $s_1\in S$, the first-stage posterior
\[
  q^\varepsilon_{s_1}:=\mu_B(\sigma^{1,\varepsilon},p)(\cdot\mid s_1)
\]
is well-defined. Moreover, for each $\theta\in\Theta$,
\[
  q^\varepsilon_{s_1}(\theta)
  =
  \frac{p(\theta)\sigma^{1,\varepsilon}_\theta(s_1)}
  {\pi^\varepsilon(s_1)}
  >0,
\]
so $q^\varepsilon_{s_1}$ has full support.

Now define
\[
  \pi^\varepsilon(s_1,s_2)
  :=
  \sum_{\theta\in\Theta}p(\theta)\sigma^{1,\varepsilon}_\theta(s_1)\sigma^2_{s_1,\theta}(s_2)
  \qquad \text{for each } (s_1,s_2)\in S\times S.
\]
For each $(s_1,s_2)$ with $\pi^\varepsilon(s_1,s_2)>0$, define
\[
  r^\varepsilon_{s_1,s_2}
  :=
  \mu_B(\sigma^2_{s_1},q^\varepsilon_{s_1})(\cdot\mid s_2)
\]
and
\[
  r^{\prime,\varepsilon}_{s_1,s_2}
  :=
  D_{D_p(q^\varepsilon_{s_1})}\!\left(
  \mu_B(\sigma^2_{s_1},D_p(q^\varepsilon_{s_1}))(\cdot\mid s_2)
  \right).
\]
These objects are well-defined whenever $\pi^\varepsilon(s_1,s_2)>0$. Indeed,
\[
  \pi^\varepsilon(s_1,s_2)
  =
  \pi^\varepsilon(s_1)
  \sum_{\theta\in\Theta}q^\varepsilon_{s_1}(\theta)\sigma^2_{s_1,\theta}(s_2),
\]
so $\pi^\varepsilon(s_1,s_2)>0$ implies
\[
  \sum_{\theta\in\Theta}q^\varepsilon_{s_1}(\theta)\sigma^2_{s_1,\theta}(s_2)>0.
\]
Since $D$ is regular, it has non-decreasing support, so
\[
  \Theta_{q^\varepsilon_{s_1}}\subseteq \Theta_{D_p(q^\varepsilon_{s_1})}.
\]
Therefore,
\[
  \sum_{\theta\in\Theta}D_p(q^\varepsilon_{s_1})(\theta)\sigma^2_{s_1,\theta}(s_2)>0,
\]
and hence $\mu_B(\sigma^2_{s_1},D_p(q^\varepsilon_{s_1}))(\cdot\mid s_2)$ is well-defined.

When $\pi^\varepsilon(s_1,s_2)=0$, assign arbitrary values to $r^\varepsilon_{s_1,s_2}$ and $r^{\prime,\varepsilon}_{s_1,s_2}$; these terms do not affect the sum below. Thus, for every bounded continuous function $f:\Delta(\Theta)\times\Delta(\Theta)\to\mathbb{R}$,
\[
  \int f\,d\tau^\varepsilon
  =
  \sum_{(s_1,s_2)\in S\times S}
  \pi^\varepsilon(s_1,s_2)\,
  f(r^\varepsilon_{s_1,s_2},r^{\prime,\varepsilon}_{s_1,s_2}).
\]

We now show that $\tau^\varepsilon\Rightarrow\tau$ as $\varepsilon\downarrow 0$. First, for every $s_1\in S$,
\[
  \pi^\varepsilon(s_1)\to \pi(s_1),
\]
and for every $(s_1,s_2)\in S\times S$,
\[
  \pi^\varepsilon(s_1,s_2)\to \pi(s_1,s_2),
\]
because $\sigma^{1,\varepsilon}\to \sigma^1$ pointwise.

Next fix $(s_1,s_2)\in S_2^+$. Then $\pi(s_1,s_2)>0$, and hence $\pi(s_1)>0$. Since
\[
  q^\varepsilon_{s_1}(\theta)
  =
  \frac{p(\theta)\bigl((1-\varepsilon)\sigma^1_\theta(s_1)+\varepsilon m(s_1)\bigr)}
  {(1-\varepsilon)\pi(s_1)+\varepsilon m(s_1)},
\]
we have $q^\varepsilon_{s_1}\to q_{s_1}$ as $\varepsilon\downarrow 0$.


Now define
\[
  \alpha_{s_1,s_2}(q):=\sum_{\theta\in\Theta}q(\theta)\sigma^2_{s_1,\theta}(s_2).
\]
Since $(s_1,s_2)\in S_2^+$,
\[
  \alpha_{s_1,s_2}(q_{s_1})
  =
  \sum_{\theta\in\Theta}
  \frac{p(\theta)\sigma_\theta^1(s_1)\sigma^2_{s_1,\theta}(s_2)}
  {\pi(s_1)}
  =
  \frac{\pi(s_1,s_2)}{\pi(s_1)}
  >0.
\]
Hence, for all sufficiently small $\varepsilon$, $\alpha_{s_1,s_2}(q^\varepsilon_{s_1})>0$, so
\[
  r^\varepsilon_{s_1,s_2}
  =
  \left(
  \frac{q^\varepsilon_{s_1}(\theta)\sigma^2_{s_1,\theta}(s_2)}
  {\alpha_{s_1,s_2}(q^\varepsilon_{s_1})}
  \right)_{\theta\in\Theta}
\]
is well-defined. By continuity of Bayes' rule on the domain where the denominator is strictly positive,
\[
  r^\varepsilon_{s_1,s_2}\to r_{s_1,s_2}.
\]

Similarly, let
\[
  \beta_{s_1,s_2}(q):=
  \sum_{\theta\in\Theta}D_p(q)(\theta)\sigma^2_{s_1,\theta}(s_2).
\]
Since $(s_1,s_2)\in S_2^+$, there exists $\theta\in\Theta$ such that $q_{s_1}(\theta)>0$ and $\sigma^2_{s_1,\theta}(s_2)>0$. By non-decreasing support, $D_p(q_{s_1})(\theta)>0$ for such $\theta$. Therefore, $\beta_{s_1,s_2}(q_{s_1})>0$. Since $D_p$ is continuous by regularity, for all sufficiently small $\varepsilon$,
\[
  \beta_{s_1,s_2}(q^\varepsilon_{s_1})>0.
\]
Hence $\mu_B(\sigma^2_{s_1},D_p(q^\varepsilon_{s_1}))(\cdot\mid s_2)$ is well-defined for all sufficiently small $\varepsilon$. By continuity of Bayes' rule,
\[
  \mu_B(\sigma^2_{s_1},D_p(q^\varepsilon_{s_1}))(\cdot\mid s_2)
  \to
  \mu_B(\sigma^2_{s_1},D_p(q_{s_1}))(\cdot\mid s_2).
\]
Using the continuity of $D$, we conclude that
\[
  r^{\prime,\varepsilon}_{s_1,s_2}\to r'_{s_1,s_2}.
\]

Therefore, for each $(s_1,s_2)\in S_2^+$,
\[
  \pi^\varepsilon(s_1,s_2)\,
  f(r^\varepsilon_{s_1,s_2},r^{\prime,\varepsilon}_{s_1,s_2})
  \to
  \pi(s_1,s_2)\,f(r_{s_1,s_2},r'_{s_1,s_2}).
\]

If $(s_1,s_2)\notin S_2^+$, then $\pi(s_1,s_2)=0$. Since $\pi^\varepsilon(s_1,s_2)\to 0$,
\[
  \left|
  \pi^\varepsilon(s_1,s_2)\,
  f(r^\varepsilon_{s_1,s_2},r^{\prime,\varepsilon}_{s_1,s_2})
  \right|
  \le
  \|f\|_\infty\,\pi^\varepsilon(s_1,s_2)\to 0.
\]

Because $S\times S$ is finite, summing over all pairs $(s_1,s_2)$ yields
\[
  \int f\,d\tau^\varepsilon \to \int f\,d\tau
\]
for every bounded continuous $f$. Hence $\tau^\varepsilon$ converges weakly to $\tau$ as $\varepsilon\downarrow 0$. Since each $\tau^\varepsilon$ belongs to $T^2_0(p,D)$, this proves that $T^2_0(p,D)$ is dense in $T^2(p,D)$ under the weak topology.
\end{proof}
\bigskip

  For the following results, we rewrite $T^1(p,D)$ and $T^2_0(p,D)$ in terms of Bayes-plausible distributions over posteriors. Let
\[
  \mathcal{R}(p):= \{\rho \in \Delta(\Delta(\Theta)) : \mathbb{E}_{\tilde{q}\sim\rho}[\tilde{q}] = p\}
\]
denote the set of Bayes-plausible distributions over posteriors given prior $p$.

Given any distortion rule $D$ and belief $p\in \Delta(\Theta)$, define $D^*_p:\Delta(\Theta)\to\Delta(\Theta)\times \Delta(\Theta)$ by
\[
  D^*_p(q) = (q, D_p(q)).
\]
Let $K$ denote the set of transition kernels $\nu:\Delta(\Theta)\to \Delta(\Delta(\Theta))$ such that $\nu(q)\in \mathcal{R}(q)$ for every $q\in \Delta(\Theta)$. Given $\rho \in \mathcal{R}(p)$ and $\nu \in K$, let $\rho\otimes \nu$ denote the measure on $\Delta(\Theta)\times\Delta(\Theta)$ induced by $\rho$ and $\nu$: for any Borel subsets $A,B$ of $\Delta(\Theta)$,
\[
  \rho\otimes \nu(A\times B) = \int_A \nu(q)(B)\, d\rho(q).
\]
Finally, given any full-support belief $p\in \Delta^\circ(\Theta)$ and distortion rule $D$, define $\bar D^{II}(\cdot;p):\Delta^\circ(\Theta)\times\Delta(\Theta)\to\Delta(\Theta)\times\Delta(\Theta)$ by
\[
  \bar{D}^{II}(q,r;p) = (r,D^{II}_q(r;p)).
\]
  
  Given Lemma \ref{lem:hetero_transformation}, we obtain the following representation of $T^1(p,D)$ and $T_0^2(p,D)$, analogous to Proposition 1 of \cite{kamenica2011bayesian}.

\begin{lem}\label{lem:representation_T1_T20}
  Consider any finite set $\Theta$, full-support prior $p\in \Delta^\circ(\Theta)$, and regular receiver distortion rule $D$. Then
  \[
    T^1(p,D)= \left\{ \tau\in\Delta(\Delta(\Theta)\times\Delta(\Theta)) :
    \text{there exists } \rho \in \mathcal{R}(p) \text{ such that } \tau = \rho \circ (D^{*}_p)^{-1} \right\},
  \]
  where $\rho\circ (D_p^*)^{-1}$ denotes the pushforward of $\rho$ under $D_p^*$, and
  \[
    \begin{aligned}
      T^2_0(p,D) = &\left\{ \tau\in\Delta(\Delta(\Theta)\times\Delta(\Theta)) :
    \text{there exist } \rho \in \mathcal{R}_0(p) \text{ and } \nu\in K \text{ such that }\right.\\
    &\left.\tau = (\rho \otimes \nu) \circ \bar{D}^{II}(\cdot;p)^{-1} \right\},
    \end{aligned}
  \]
  where $\mathcal{R}_0(p):=\{\rho\in \mathcal{R}(p): \rho(\Delta^\circ(\Theta)) = 1\}$ and $(\rho \otimes \nu) \circ \bar{D}^{II}(\cdot;p)^{-1}$ denotes the pushforward of $\rho \otimes \nu$ under $\bar{D}^{II}(\cdot;p)$.
\end{lem}


  Using this representation, we establish the following properties of these sets.

\begin{lem}\label{lem:T1_nonempty_compact_convex}
  Consider any finite set $\Theta$, full-support prior $p\in \Delta^\circ(\Theta)$, and regular receiver distortion rule $D$. Then $T^1(p,D)$ is nonempty, compact, and convex under the topology of weak convergence.
\end{lem}
\begin{proof}
  
Nonemptiness follows because, for example, $\tau=\delta_{(p,D_p(p))}$ belongs to $T^1(p,D)$ by taking $\rho=\delta_p\in\mathcal R(p)$.

To show compactness, define $F:\mathcal{R}(p)\to\Delta(\Delta(\Theta)\times\Delta(\Theta))$ by
\[
  F(\rho):=\rho\circ (D_p^*)^{-1}.
\]
Because $D$ is regular, $D_p^*$ is continuous, and therefore the induced pushforward map $F$ is continuous under the topology of weak convergence. Since $T^1(p,D)=F(\mathcal{R}(p))$ and $\mathcal{R}(p)$ is compact, $T^1(p,D)$ is compact.

Finally, we show convexity. Let $\tau_1,\tau_2\in T^1(p,D)$ and $\lambda\in[0,1]$. Then there exist $\rho_1,\rho_2\in\mathcal R(p)$ such that
\[
  \tau_i=\rho_i\circ (D_p^*)^{-1}, \qquad i=1,2.
\]
Hence, for every Borel set $B\subseteq \Delta(\Theta)\times\Delta(\Theta)$,
\[
\begin{aligned}
  (\lambda\tau_1+(1-\lambda)\tau_2)(B)
  &=\lambda\tau_1(B)+(1-\lambda)\tau_2(B)\\
  &=\lambda\int \mathbbm{1}_{\{D_p^*(q)\in B\}}\,d\rho_1(q)
    +(1-\lambda)\int \mathbbm{1}_{\{D_p^*(q)\in B\}}\,d\rho_2(q)\\
  &=\int \mathbbm{1}_{\{D_p^*(q)\in B\}}\,d\bigl(\lambda\rho_1+(1-\lambda)\rho_2\bigr)(q).
\end{aligned}
\]
Since $\mathcal R(p)$ is convex, $\lambda\rho_1+(1-\lambda)\rho_2\in\mathcal R(p)$. Therefore,
\[
  \lambda\tau_1+(1-\lambda)\tau_2
  =
  \bigl(\lambda\rho_1+(1-\lambda)\rho_2\bigr)\circ (D_p^*)^{-1}
  \in T^1(p,D).
\]
Thus $T^1(p,D)$ is convex.
\end{proof}

\begin{lem}\label{lem:divisible_then_T1_equal_T20}
  If $D$ is a regular and divisible distortion rule, then $T^1(p,D)=T_0^2(p,D)$ for every finite set $\Theta$ and full-support prior $p\in \Delta^\circ(\Theta)$.
\end{lem}

\begin{proof}
  
Suppose $D$ is regular and divisible. By regularity, both $D_p$ and $D_q^{II}(\cdot;p)$ are continuous. Since $\Delta^\circ(\Theta)$ is dense in $\Delta(\Theta)$, divisibility and continuity imply that
\[
  D_q^{II}(r;p)=D_p(r)
\]
for all $p,q\in\Delta^\circ(\Theta)$ and $r\in\Delta(\Theta)$.

We first show that $T^1(p,D)\subseteq T^2_0(p,D)$. Take any $\tau\in T^1(p,D)$. By Lemma \ref{lem:representation_T1_T20}, there exists $\rho\in\mathcal R(p)$ such that
\[
  \tau=\rho\circ(D_p^*)^{-1}.
\]
Let $\rho':=\delta_p\in\mathcal R_0(p)$, and choose $\nu\in K$ such that $\nu(p)=\rho$. Such a kernel exists; for example, set $\nu(p)=\rho$ and $\nu(q)=\delta_q$ for all $q\neq p$. Since $D_q^{II}(r;p)=D_p(r)$ for all $p,q\in\Delta^\circ(\Theta)$ and $r\in\Delta(\Theta)$,
\[
  (\rho'\otimes\nu)\circ\bar D^{II}(\cdot;p)^{-1}
  =
  \rho\circ(D_p^*)^{-1}
  =
  \tau.
\]
Hence $\tau\in T^2_0(p,D)$ by Lemma \ref{lem:representation_T1_T20}. Therefore, $T^1(p,D)\subseteq T^2_0(p,D)$.

Next, we show that $T^2_0(p,D)\subseteq T^1(p,D)$. Take any $\tau\in T^2_0(p,D)$. By Lemma \ref{lem:representation_T1_T20}, there exist $\rho\in\mathcal R_0(p)$ and $\nu\in K$ such that
\[
  \tau=(\rho\otimes\nu)\circ\bar D^{II}(\cdot;p)^{-1}.
\]
Since $D_q^{II}(r;p)=D_p(r)$ for all $p,q\in\Delta^\circ(\Theta)$ and $r\in\Delta(\Theta)$, we have
\[
  \bar D^{II}(q,r;p)=(r,D_p(r))
\]
for $(\rho\otimes\nu)$-almost every $(q,r)$.

Let $\operatorname{pr}_2:\Delta(\Theta)\times\Delta(\Theta)\to\Delta(\Theta)$ denote the projection onto the second coordinate, and define
\[
  \rho^*:=(\rho\otimes\nu)\circ\operatorname{pr}_2^{-1}.
\]
Then
\[
  \tau=\rho^*\circ(D_p^*)^{-1}.
\]
Moreover, $\rho^*\in\mathcal R(p)$ by the law of iterated expectations, because $\rho\in\mathcal R_0(p)$ and $\nu(q)\in\mathcal R(q)$ for every $q$. Therefore, Lemma \ref{lem:representation_T1_T20} implies that $\tau\in T^1(p,D)$.

Thus $T^2_0(p,D)\subseteq T^1(p,D)$, and hence $T^1(p,D)=T^2_0(p,D)$.
\end{proof}

  \vspace{\baselineskip}
  
  In the following proof, it is convenient to focus on environments with state-independent sender preferences. In such environments, with a slight abuse of notation, let $v:A\to \mathbb{R}$ denote the sender's utility function. The sender's indirect utility is then
\[
  \hat{v}(q') := v(\hat{a}(q')).
\]

  In general, determining the feasible sets of the sender's ex ante payoffs requires the sets of joint distributions of the sender's and receiver's posterior beliefs, $T^1(p,D)$, $T^2(p,D)$, and $T^2_0(p,D)$. With state-independent sender preferences, however, only the distributions of the receiver's posterior beliefs are relevant. Define
\[
\begin{aligned}
\bar{T}^1(p,D):=\Bigl\{\tau\in\Delta(\Delta(\Theta)) :
&\ \exists\sigma\in \Sigma \text{ such that, for all } q',\\
&\ \tau(q')=\sum_{\theta\in\Theta}p(\theta)
\sum_{s\in S^1(q';\sigma,p,D)}\sigma_\theta(s)
\Bigr\},
\end{aligned}
\]
where
\[
  S^1(q';\sigma,p,D)
  :=
  \{s\in S:D_p(\mu_B(\sigma,p)(\cdot\mid s)) = q'\}.
\]
Similarly,
\[
\begin{aligned}
\bar{T}^2(p,D):=\Bigl\{&\tau\in\Delta(\Delta(\Theta)) :
\ \exists\sigma^1\in\Sigma,\ \{\sigma^2_{s_1}\}_{s_1\in S}\in\Sigma^S
\text{ such that, for all } r',\\
&\ \tau(r')=\sum_{\theta\in\Theta}p(\theta)
\sum_{s_1\in S}\sigma^1_\theta(s_1)
\sum_{s_2\in S^2(r';s_1,\sigma^1,\{\sigma^2_{s_1}\}_{s_1\in S},p,D)}
\sigma^2_{s_1,\theta}(s_2)
\Bigr\},
\end{aligned}
\]
where
\[
\begin{aligned}
&S^2(r';s_1,\sigma^1,\{\sigma^2_{s_1}\}_{s_1\in S},p,D) \\
&\quad :=
\Bigl\{s_2\in S:
r' = D_{D_p(q)}\bigl(\mu_B(\sigma^2_{s_1},D_p(q))(\cdot\mid s_2)\bigr),\
q = \mu_B(\sigma^1,p)(\cdot\mid s_1)
\Bigr\}.
\end{aligned}
\]
Finally, define the restricted set
\[
\begin{aligned}
\bar{T}^2_0(p,D):=\Bigl\{&\tau\in\Delta(\Delta(\Theta)) :
\ \exists\sigma^1\in\Sigma,\ \{\sigma^2_{s_1}\}_{s_1\in S}\in\Sigma^S
\text{ such that} \\
&\ \text{for every }s_1\in S,\ \text{if } \sum_{\theta\in\Theta}p(\theta)\sigma^1_\theta(s_1)>0,
\text{ then } \sigma^1_\theta(s_1)>0 \text{ for every }\theta\in\Theta_p,\\
&\ \text{and, for all } r',\\
&\ \tau(r')=\sum_{\theta\in\Theta}p(\theta)
\sum_{s_1\in S}\sigma^1_\theta(s_1)
\sum_{s_2\in S^2(r';s_1,\sigma^1,\{\sigma^2_{s_1}\}_{s_1\in S},p,D)}
\sigma^2_{s_1,\theta}(s_2)
\Bigr\}.
\end{aligned}
\]

  We have the following characterization of these sets, analogous to Lemma \ref{lem:representation_T1_T20}.

\begin{lem}\label{lem:representation_T1_T20_state_independent}
  Consider any finite set $\Theta$, full-support prior $p\in \Delta^\circ(\Theta)$, and regular receiver distortion rule $D$. Then
  \[
    \bar{T}^1(p,D)= \left\{ \tau\in\Delta(\Delta(\Theta)) :
    \text{there exists } \rho \in \mathcal{R}(p) \text{ such that } \tau = \rho \circ (D_p)^{-1} \right\},
  \]
  and
  \[
    \begin{aligned}
      \bar{T}^2_0(p,D) = \Bigl\{& \tau\in\Delta(\Delta(\Theta)) :
    \text{there exist } \rho \in \mathcal{R}_0(p) \text{ and } \nu\in K \text{ such that }\\
    &\tau = (\rho \otimes \nu) \circ D^{II}(\cdot;p)^{-1} \Bigr\},
    \end{aligned}
  \]
  where, with a slight abuse of notation, $D^{II}(\cdot;p):\Delta^\circ(\Theta)\times\Delta(\Theta)\to\Delta(\Theta)$ is defined by
  \[
    D^{II}(q,r;p) = D^{II}_q(r;p).
  \]
\end{lem}

  
  We use the following properties of these sets.

\begin{lem}\label{lem:bar_T1_nonempty_compact_convex}
  Consider any finite set $\Theta$, full-support prior $p\in \Delta^\circ(\Theta)$, and regular receiver distortion rule $D$. Then $\bar{T}^1(p,D)$ is nonempty, compact, and convex under the topology of weak convergence.
\end{lem}

\begin{proof}
  The proof is analogous to that of Lemma \ref{lem:T1_nonempty_compact_convex}.
\end{proof}

\begin{lem}\label{lem:bar_T20_dense_in_bar_T2}
  Consider any finite set $\Theta$, full-support prior $p\in \Delta^\circ(\Theta)$, and regular receiver distortion rule $D$. Then $\bar{T}_0^2(p,D)$ is dense in $\bar{T}^2(p,D)$ under the topology of weak convergence.
\end{lem}

\begin{proof}
  The proof is analogous to that of Lemma \ref{lem:T20_dense_in_T2}.
\end{proof}


  We also use the following properties related to divisibility.

\begin{lem}\label{lem:bar_T20_sub_bar_T1_then_matrix_divisible}
  Consider any finite set $\Theta$, full-support prior $p\in \Delta^\circ(\Theta)$, and regular receiver distortion rule $D$. If $\bar{T}_0^2(p,D)\subseteq \bar{T}^1(p,D)$, then for every $q\in \Delta^\circ(\Theta)$, there exists a full-rank $|\Theta|\times |\Theta|$ column-stochastic matrix $M_{p,q}$ such that
  \[
    D_p(M_{p,q} r) = D^{II}_q(r;p) \quad \text{for every } r\in \Delta(\Theta).
  \]
\end{lem}
\begin{proof}
  
By Lemma \ref{lem:representation_T1_T20_state_independent}, the inclusion
\[
  \bar{T}_0^2(p,D)\subseteq \bar{T}^1(p,D)
\]
implies that, for every $\rho\in\mathcal R_0(p)$ and $\nu\in K$, there exists $\tilde\rho\in\mathcal R(p)$ such that
\[
  (\rho\otimes\nu)\circ D^{II}(\cdot;p)^{-1}
  =
  \tilde\rho\circ (D_p)^{-1},
\]
where $(D_p)^{-1}$ denotes the inverse image map used in the pushforward. Since $\tilde\rho\circ(D_p)^{-1}$ is supported on $D_p(\Delta(\Theta))$, we have
\[
  (\rho\otimes\nu)
  \left(
    \left\{(q,r):D_q^{II}(r;p)\in D_p(\Delta(\Theta))\right\}
  \right)
  =1.
\]
Thus, for $(\rho\otimes\nu)$-almost every $(q,r)$,
\[
  D_p^{-1}\bigl(D_q^{II}(r;p)\bigr)
\]
is well-defined, where $D_p^{-1}$ denotes the inverse function on the image of $D_p$.

We first show that this expression is well-defined for every $q\in\Delta^\circ(\Theta)$ and $r\in\Delta(\Theta)$. Suppose not. Then there exist $q^*\in\Delta^\circ(\Theta)$ and $r^*\in\Delta(\Theta)$ such that
\[
  D_p^{-1}\bigl(D_{q^*}^{II}(r^*;p)\bigr)
\]
is not well-defined. Since $p,q^*\in\Delta^\circ(\Theta)$, there exist $\alpha\in(0,1)$ and $q^\dagger\in\Delta^\circ(\Theta)$ such that
\[
  p=\alpha q^*+(1-\alpha)q^\dagger.
\]
Hence
\[
  \rho:=\alpha\delta_{q^*}+(1-\alpha)\delta_{q^\dagger}\in\mathcal R_0(p).
\]
Similarly, because $q^*$ has full support, there exist $\beta\in(0,1)$ and $r^\dagger\in\Delta(\Theta)$ such that
\[
  q^*=\beta r^*+(1-\beta)r^\dagger.
\]
Define
\[
  \nu(q^*):=\beta\delta_{r^*}+(1-\beta)\delta_{r^\dagger},
\]
and set $\nu(\bar q):=\delta_{\bar q}$ for all $\bar q\neq q^*$. Then $\nu\in K$. Moreover,
\[
  (\rho\otimes\nu)(\{(q^*,r^*)\})=\alpha\beta>0,
\]
contradicting the almost-sure well-definedness above. Therefore,
\[
  D_p^{-1}\bigl(D_q^{II}(r;p)\bigr)
\]
is well-defined for every $q\in\Delta^\circ(\Theta)$ and $r\in\Delta(\Theta)$.

Since $\tilde\rho\in\mathcal R(p)$, the equality of pushforward measures implies
\[
  \int\!\!\int D_p^{-1}\bigl(D_q^{II}(r;p)\bigr)\,d\nu(q)(r)\,d\rho(q)
  =
  p.
\]
On the other hand, since $\rho\in\mathcal R_0(p)$ and $\nu(q)\in\mathcal R(q)$ for every $q$,
\[
  \int\!\!\int r\,d\nu(q)(r)\,d\rho(q)
  =
  \int q\,d\rho(q)
  =
  p.
\]
Therefore,
\begin{equation}\label{eqn:monent_condition}
  \int\!\!\int
  \left[
    D_p^{-1}\bigl(D_q^{II}(r;p)\bigr)-r
  \right]
  d\nu(q)(r)\,d\rho(q)
  =
  0.
\end{equation}

Define
\[
  f(r;p,q):=D_p^{-1}\bigl(D_q^{II}(r;p)\bigr)-r
\]
for $p,q\in\Delta^\circ(\Theta)$ and $r\in\Delta(\Theta)$. We show that $f(\cdot;p,q)$ is affine. For any $\rho_q\in\mathcal R(q)$, define
\[
  H(q,\rho_q):=\int_{\Delta(\Theta)} f(r;p,q)\,d\rho_q(r).
\]
We first show that $H(q,\cdot)$ is constant on $\mathcal R(q)$ for every $q\in\Delta^\circ(\Theta)$.

Suppose not. Then there exist $q^*\in\Delta^\circ(\Theta)$ and $\rho_q^1,\rho_q^2\in\mathcal R(q^*)$ such that
\[
  H(q^*,\rho_q^1)\neq H(q^*,\rho_q^2).
\]
Since $p,q^*\in\Delta^\circ(\Theta)$, there exist $\lambda\in(0,1)$ and $q^\dagger\in\Delta^\circ(\Theta)$ such that
\[
  p=\lambda q^*+(1-\lambda)q^\dagger.
\]
Let
\[
  \rho^*:=\lambda\delta_{q^*}+(1-\lambda)\delta_{q^\dagger}\in\mathcal R_0(p).
\]
Choose kernels $\nu^1,\nu^2\in K$ such that
\[
  \nu^1(q^*)=\rho_q^1,\qquad
  \nu^2(q^*)=\rho_q^2,
\]
and $\nu^1(q)=\nu^2(q)$ for all $q\neq q^*$. Applying \eqref{eqn:monent_condition} to $(\rho^*,\nu^1)$ and $(\rho^*,\nu^2)$ yields
\[
  \lambda H(q^*,\rho_q^1)+(1-\lambda)H(q^\dagger,\nu^1(q^\dagger))
  =
  0
  =
  \lambda H(q^*,\rho_q^2)+(1-\lambda)H(q^\dagger,\nu^2(q^\dagger)).
\]
Since $\nu^1(q^\dagger)=\nu^2(q^\dagger)$, this implies
\[
  H(q^*,\rho_q^1)=H(q^*,\rho_q^2),
\]
a contradiction. Hence $H(q,\cdot)$ is constant on $\mathcal R(q)$ for every $q\in\Delta^\circ(\Theta)$.

We now prove that $f(\cdot;p,q)$ is affine. Fix $q\in\Delta^\circ(\Theta)$, $r_1,r_2\in\Delta(\Theta)$, and $\alpha\in[0,1]$. Let
\[
  \bar r:=\alpha r_1+(1-\alpha)r_2.
\]
Because $q$ has full support, for sufficiently small $\varepsilon\in(0,1)$ there exists $r^\dagger\in\Delta(\Theta)$ such that
\[
  q=\varepsilon \bar r+(1-\varepsilon)r^\dagger.
\]
Define two measures in $\mathcal R(q)$ by
\[
  \rho_q^1
  :=
  \varepsilon\alpha\delta_{r_1}
  +\varepsilon(1-\alpha)\delta_{r_2}
  +(1-\varepsilon)\delta_{r^\dagger}
\]
and
\[
  \rho_q^2
  :=
  \varepsilon\delta_{\bar r}
  +(1-\varepsilon)\delta_{r^\dagger}.
\]
Since $H(q,\cdot)$ is constant on $\mathcal R(q)$,
\[
  \int f(r;p,q)\,d\rho_q^1(r)
  =
  \int f(r;p,q)\,d\rho_q^2(r).
\]
Canceling the common term $(1-\varepsilon)f(r^\dagger;p,q)$ and dividing by $\varepsilon$ gives
\[
  \alpha f(r_1;p,q)+(1-\alpha)f(r_2;p,q)
  =
  f(\alpha r_1+(1-\alpha)r_2;p,q).
\]
Thus $f(\cdot;p,q)$ is affine.

Hence there exist a $|\Theta|\times|\Theta|$ matrix $X_{p,q}$ and a vector $k_{p,q}\in\mathbb R^{|\Theta|}$ such that
\[
  f(r;p,q)=X_{p,q}r+k_{p,q}
  \qquad\text{for all } r\in\Delta(\Theta).
\]
Let $X'_{p,q}:=X_{p,q}+I$. Then
\[
  D_p^{-1}\bigl(D_q^{II}(r;p)\bigr)
  =
  X'_{p,q}r+k_{p,q}
  \qquad\text{for all } r\in\Delta(\Theta).
\]
Since $\sum_{\theta\in\Theta}r(\theta)=1$ for all $r\in\Delta(\Theta)$, define
\[
  M_{p,q}:=X'_{p,q}+k_{p,q}\mathbf 1^\top,
\]
where $\mathbf 1$ is the $|\Theta|$-dimensional vector of ones. Then
\[
  D_p^{-1}\bigl(D_q^{II}(r;p)\bigr)=M_{p,q}r
  \qquad\text{for all } r\in\Delta(\Theta).
\]
Equivalently,
\[
  D_p(M_{p,q}r)=D_q^{II}(r;p)
  \qquad\text{for all } r\in\Delta(\Theta).
\]

For each $\theta\in\Theta$, $M_{p,q}\delta_\theta\in\Delta(\Theta)$, and this vector is the $\theta$-th column of $M_{p,q}$. Hence each column of $M_{p,q}$ is a probability vector, so $M_{p,q}$ is column-stochastic.

Finally, $D_p^{-1}\circ D_q^{II}(\cdot;p)$ is injective. Indeed, $D_q^{II}(\cdot;p)$ is injective because it is the composition of the injective map
\[
  r\mapsto \mathcal B\left(\frac{D_p(q)}{q},r\right)
\]
and the injective map $D_{D_p(q)}$, and $D_p^{-1}$ is injective on $D_p(\Delta(\Theta))$. Thus $M_{p,q}$ is injective on $\Delta(\Theta)$. Since $M_{p,q}$ is column-stochastic, if it were not full rank, there would exist a nonzero vector $z$ with $M_{p,q}z=0$. Column-stochasticity implies $\mathbf 1^\top z=0$. For any interior belief $r\in\Delta^\circ(\Theta)$ and sufficiently small $t>0$, both $r+tz$ and $r-tz$ belong to $\Delta(\Theta)$ and are distinct, but
\[
  M_{p,q}(r+tz)=M_{p,q}(r-tz),
\]
contradicting injectivity on $\Delta(\Theta)$. Therefore, $M_{p,q}$ is full rank.
\end{proof}

\begin{lem}\label{lem:weakly_divisible_to_bar_T20_sub_bar_T1}
  Suppose $D$ is regular. Consider any finite set $\Theta$ and full-support prior $p$. If $D$ is weakly divisible, then $\bar{T}^2(p,D)=\bar{T}^1(p,D)$.
\end{lem}

\begin{proof}
  
We first show that $\bar{T}_0^2(p,D)\subseteq \bar{T}^1(p,D)$.

By weak divisibility, for every $q\in\Delta^\circ(\Theta)$, there exists a full-rank column-stochastic matrix $M_{p,q}$ such that
\[
  D_p(M_{p,q}r)=D_q^{II}(r;p)
  \quad\text{for every } r\in\Delta(\Theta).
\]
Define $M(q,r):=M_{p,q}r$ for $q\in\Delta^\circ(\Theta)$ and $r\in\Delta(\Theta)$.\footnote{The map can be chosen measurably because $M_{p,q}r=D_p^{-1}(D_q^{II}(r;p))$, and the right-hand side is continuous in $(q,r)$ on $\Delta^\circ(\Theta)\times\Delta(\Theta)$.}

Take any $\tau\in\bar{T}_0^2(p,D)$. By Lemma \ref{lem:representation_T1_T20_state_independent}, there exist $\rho\in\mathcal R_0(p)$ and $\nu\in K$ such that
\[
  \tau=(\rho\otimes\nu)\circ D^{II}(\cdot;p)^{-1}.
\]
Define $\tilde\rho\in\Delta(\Delta(\Theta))$ by
\[
  \tilde\rho(B)
  :=
  \int_{\Delta^\circ(\Theta)}
  \nu(q)\left(\{r\in\Delta(\Theta):M_{p,q}r\in B\}\right)
  d\rho(q)
\]
for every Borel set $B\subseteq\Delta(\Theta)$. By weak divisibility, for every Borel set $B\subseteq\Delta(\Theta)$,
\[
\begin{aligned}
  \tau(B)
  &=
  (\rho\otimes\nu)\left(\{(q,r):D_q^{II}(r;p)\in B\}\right)\\
  &=
  (\rho\otimes\nu)\left(\{(q,r):D_p(M_{p,q}r)\in B\}\right)\\
  &=
  \tilde\rho\bigl(D_p^{-1}(B)\bigr)\\
  &=
  \tilde\rho\circ(D_p)^{-1}(B).
\end{aligned}
\]
Thus,
\[
  \tau=\tilde\rho\circ(D_p)^{-1}.
\]
To show that $\tau\in\bar{T}^1(p,D)$, it remains to verify that $\tilde\rho\in\mathcal R(p)$.

Because $\nu(q)\in\mathcal R(q)$ for every $q$, we have
\[
\begin{aligned}
  \int_{\Delta(\Theta)} x\,d\tilde\rho(x)
  &=
  \int_{\Delta^\circ(\Theta)}
  \int_{\Delta(\Theta)} M_{p,q}r\,d\nu(q)(r)\,d\rho(q)\\
  &=
  \int_{\Delta^\circ(\Theta)}
  M_{p,q}\left(\int_{\Delta(\Theta)} r\,d\nu(q)(r)\right)d\rho(q)\\
  &=
  \int_{\Delta^\circ(\Theta)} M_{p,q}q\,d\rho(q).
\end{aligned}
\]
By invariance under no information, $D_x(x)=x$ for every $x\in\Delta^\circ(\Theta)$. Hence, for every $p,q\in\Delta^\circ(\Theta)$,
\[
  D_q^{II}(q;p)
  =
  D_{D_p(q)}(D_p(q))
  =
  D_p(q).
\]
Weak divisibility therefore gives
\[
  D_p(M_{p,q}q)=D_p(q).
\]
Since $D_p$ is one-to-one by regularity,
\[
  M_{p,q}q=q.
\]
Thus,
\[
  \int_{\Delta(\Theta)} x\,d\tilde\rho(x)
  =
  \int_{\Delta^\circ(\Theta)} q\,d\rho(q)
  =
  p,
\]
because $\rho\in\mathcal R_0(p)\subseteq\mathcal R(p)$. Hence $\tilde\rho\in\mathcal R(p)$, and Lemma \ref{lem:representation_T1_T20_state_independent} implies $\tau\in\bar{T}^1(p,D)$. Therefore,
\[
  \bar{T}_0^2(p,D)\subseteq \bar{T}^1(p,D).
\]

Next, we show that $\bar{T}^1(p,D)\subseteq \bar{T}_0^2(p,D)$. Take any $\tau\in\bar{T}^1(p,D)$. By Lemma \ref{lem:representation_T1_T20_state_independent}, there exists $\rho\in\mathcal R(p)$ such that
\[
  \tau=\rho\circ(D_p)^{-1}.
\]
Let $\rho':=\delta_p\in\mathcal R_0(p)$, and choose $\nu\in K$ such that $\nu(p)=\rho$; for example, set $\nu(p)=\rho$ and $\nu(q)=\delta_q$ for all $q\neq p$.

By invariance under no information, $D_p(p)=p$. Therefore,
\[
  D_p^{II}(r;p)
  =
  D_{D_p(p)}
  \left(
    \mathcal B\left(\frac{D_p(p)}{p},r\right)
  \right)
  =
  D_p(r).
\]
Hence,
\[
  (\rho'\otimes\nu)\circ D^{II}(\cdot;p)^{-1}
  =
  \rho\circ(D_p)^{-1}
  =
  \tau.
\]
By Lemma \ref{lem:representation_T1_T20_state_independent}, this implies $\tau\in\bar{T}_0^2(p,D)$. Therefore,
\[
  \bar{T}^1(p,D)\subseteq\bar{T}_0^2(p,D).
\]

We have shown
\[
  \bar{T}^1(p,D)=\bar{T}_0^2(p,D).
\]
By Lemma \ref{lem:bar_T20_dense_in_bar_T2}, $\bar{T}_0^2(p,D)$ is dense in $\bar{T}^2(p,D)$, and by Lemma \ref{lem:bar_T1_nonempty_compact_convex}, $\bar{T}^1(p,D)$ is closed. Since $\bar{T}^1(p,D)=\bar{T}_0^2(p,D)\subseteq\bar{T}^2(p,D)$, it follows that
\[
  \bar{T}^1(p,D)=\bar{T}^2(p,D).
\]
\end{proof}

\begin{lem}\label{lem:T20_sub_T1_then_divisible}
  Suppose $D$ is regular. If $T_0^2(p,D)\subseteq T^1(p,D)$ for every finite state space $\Theta$ and full-support common prior $p$, then $D$ is divisible.
\end{lem}

\begin{proof}
  
By Lemma \ref{lem:representation_T1_T20}, the inclusion
\[
  T_0^2(p,D)\subseteq T^1(p,D)
\]
implies that, for every $\rho\in\mathcal R_0(p)$ and $\nu\in K$, there exists $\tilde\rho\in\mathcal R(p)$ such that
\[
  (\rho\otimes\nu)\circ\bar D^{II}(\cdot;p)^{-1}
  =
  \tilde\rho\circ(D_p^*)^{-1}.
\]
Since $D_p^*(\Delta(\Theta))=\operatorname{Gr}(D_p)$, the measure on the right-hand side is supported on $\operatorname{Gr}(D_p)$. Hence,
\[
  (\rho\otimes\nu)\circ\bar D^{II}(\cdot;p)^{-1}
  \bigl(\operatorname{Gr}(D_p)\bigr)=1.
\]
Since $\bar D^{II}(q,r;p)=(r,D_q^{II}(r;p))$, this means
\[
  (\rho\otimes\nu)
  \left(
    \left\{(q,r):(r,D_q^{II}(r;p))\in\operatorname{Gr}(D_p)\right\}
  \right)=1.
\]
Equivalently,
\[
  (\rho\otimes\nu)
  \left(
    \left\{(q,r):D_q^{II}(r;p)=D_p(r)\right\}
  \right)=1.
\]
Thus,
\[
  D_q^{II}(r;p)=D_p(r)
\]
for $(\rho\otimes\nu)$-almost every $(q,r)$.

Because $\rho\in\mathcal R_0(p)$ and $\nu\in K$ can be chosen arbitrarily, this almost-sure equality implies pointwise equality on full-support beliefs. To see this, fix any $p,q,r\in\Delta^\circ(\Theta)$. For sufficiently small $\varepsilon>0$, define
\[
  x:=\frac{p-\varepsilon q}{1-\varepsilon}.
\]
Then $x\in\Delta^\circ(\Theta)$ and
\[
  p=\varepsilon q+(1-\varepsilon)x.
\]
Similarly, for sufficiently small $\eta>0$, define
\[
  y:=\frac{q-\eta r}{1-\eta}.
\]
Then $y\in\Delta^\circ(\Theta)$ and
\[
  q=\eta r+(1-\eta)y.
\]
Set
\[
  \rho:=\varepsilon\delta_q+(1-\varepsilon)\delta_x.
\]
Then $\rho\in\mathcal R_0(p)$. Define $\nu\in K$ by
\[
  \nu(q):=\eta\delta_r+(1-\eta)\delta_y,
  \qquad
  \nu(z):=\delta_z \quad\text{for every } z\neq q.
\]
Then \(\nu(q)\in\mathcal R(q)\), and hence \(\nu\in K\). Moreover,
\[
  (\rho\otimes\nu)(\{(q,r)\})=\varepsilon\eta>0.
\]
The almost-sure equality above therefore implies
\[
  D_q^{II}(r;p)=D_p(r).
\]
Since $p,q,r\in\Delta^\circ(\Theta)$ were arbitrary, $D$ is divisible.
\end{proof}


\subsection{Proof of Proposition \ref{prop:grether}}\label{app:proof_grether}

Suppose the receiver updates according to Grether's $\alpha$--$\beta$ rule for some $\alpha>0$ and $\beta>0$. We compare the values of the static and dynamic persuasion problems. As in \cite{de2022non}, the value of the static persuasion problem is
\[
  \sup_{x\in \mathcal{V}^1(p,D)}x
  =
  \left[\operatorname{CAV}(\check{v}^{D_p})\right](p)
  =
  \begin{cases}
    p\left[1+\left(\frac{p}{1-p}\right)^{\frac{\alpha}{\beta}-1}\right]
    &\text{if }p<\frac{1}{2},\\
    1
    &\text{if }p\geq \frac{1}{2}.
  \end{cases}
\]

We next derive the value of the dynamic persuasion problem. When the receiver updates according to Grether's $\alpha$--$\beta$ rule, $q$ and $D_p(q)$ have the same support for every $p\in\Delta(\Theta)$ and $q\in\Delta(\Theta_p)$. Thus, by Lemma \ref{lem:hetero_transformation}, the value of the interim problem, conditional on sender interim belief $q$, is
\[
  \sup_{\rho\in\mathcal{R}(q)}
  \mathbb{E}_{\tilde q\sim\rho}
  \left[
    \check{v}^{D^{II}_q(\cdot;p)}(\tilde q)
  \right]
  =
  \left[\operatorname{CAV}
  \left(\check{v}^{D^{II}_q(\cdot;p)}\right)
  \right](q).
\]
Denote this value by $v^{II}(q)$. The value of the dynamic persuasion problem is then
\[
  \sup_{\rho\in\mathcal{R}(p)}
  \mathbb{E}_{\tilde q\sim\rho}\left[v^{II}(\tilde q)\right]
  =
  \left[\operatorname{CAV}(v^{II})\right](p).
\]
We compare $\left[\operatorname{CAV}(v^{II})\right](p)$ with $\left[\operatorname{CAV}(\check{v}^{D_p})\right](p)$.

For any $p\in\Delta^\circ(\Theta)$, $q\in\Delta(\Theta)$, $r\in\Delta(\Theta_q)$, and $\theta\in\Theta$,
\[
  D_q^{II}(r;p)(\theta)
  =
  \frac{
    \mathbbm{1}_{\{q(\theta)>0\}}
    p(\theta)^{\alpha(\alpha-\beta)}
    q(\theta)^{(\alpha-1)\beta}
    r(\theta)^\beta
  }{
    \sum_{\theta'\in\Theta}
    \mathbbm{1}_{\{q(\theta')>0\}}
    p(\theta')^{\alpha(\alpha-\beta)}
    q(\theta')^{(\alpha-1)\beta}
    r(\theta')^\beta
  }.
\]

When $\Theta=\{0,1\}$, identify each belief with its mass on state $1$. Then
\[
  D_q^{II}(r;p)
  =
  \begin{cases}
    \dfrac{
      p^{\alpha(\alpha-\beta)}
      q^{(\alpha-1)\beta}
      r^\beta
    }{
      p^{\alpha(\alpha-\beta)}
      q^{(\alpha-1)\beta}
      r^\beta
      +
      (1-p)^{\alpha(\alpha-\beta)}
      (1-q)^{(\alpha-1)\beta}
      (1-r)^\beta
    }
    &\text{if }q\in(0,1),\\
    0
    &\text{if }q=0,\\
    1
    &\text{if }q=1.
  \end{cases}
\]

In the judge-prosecutor environment,
\[
  \check{v}^{D_q^{II}(\cdot;p)}(r)
  =
  \mathbbm{1}\left\{D_q^{II}(r;p)\geq \frac{1}{2}\right\}.
\]
Therefore,
\[
  \check{v}^{D_q^{II}(\cdot;p)}(r)
  =
  \mathbbm{1}
  \left\{
    q=1
    \text{ or }
    \left(q\in(0,1)\text{ and }r\geq r_{\alpha,\beta}(p,q)\right)
  \right\},
\]
where
\[
  r_{\alpha,\beta}(p,q)
  =
  \frac{
    (1-p)^{\frac{\alpha}{\beta}(\alpha-\beta)}
    (1-q)^{\alpha-1}
  }{
    (1-p)^{\frac{\alpha}{\beta}(\alpha-\beta)}
    (1-q)^{\alpha-1}
    +
    p^{\frac{\alpha}{\beta}(\alpha-\beta)}
    q^{\alpha-1}
  }.
\]
Thus,
\[
  \left[
    \operatorname{CAV}
    \left(\check{v}^{D_q^{II}(\cdot;p)}\right)
  \right](r)
  =
  \begin{cases}
    \dfrac{r}{r_{\alpha,\beta}(p,q)}
    &\text{if }q\in(0,1)\text{ and }r<r_{\alpha,\beta}(p,q),\\
    1
    &\text{if }q=1\text{ or }(q\in(0,1)\text{ and }r\geq r_{\alpha,\beta}(p,q)),\\
    0
    &\text{if }q=0.
  \end{cases}
\]
Hence, the value of the interim problem starting from sender posterior $q$ is
\[
\begin{aligned}
  v^{II}(q)
  &=
  \left[
    \operatorname{CAV}
    \left(\check{v}^{D_q^{II}(\cdot;p)}\right)
  \right](q)\\
  &=
  \begin{cases}
    \dfrac{q}{r_{\alpha,\beta}(p,q)}
    &\text{if }q\in(0,1)\text{ and }q<r_{\alpha,\beta}(p,q),\\
    1
    &\text{if }q=1\text{ or }(q\in(0,1)\text{ and }q\geq r_{\alpha,\beta}(p,q)),\\
    0
    &\text{if }q=0.
  \end{cases}
\end{aligned}
\]
Equivalently,
\[
  v^{II}(q)
  =
  \begin{cases}
    0
    &\text{if }q=0,\\
    q\left[
      1+
      \left(\frac{p}{1-p}\right)^{\frac{\alpha}{\beta}(\alpha-\beta)}
      \left(\frac{q}{1-q}\right)^{\alpha-1}
    \right]
    &\text{if }q\in(0,q_{\alpha,\beta}(p)),\\
    1
    &\text{if }q\in[q_{\alpha,\beta}(p),1],
  \end{cases}
\]
where
\[
  q_{\alpha,\beta}(p)
  =
  \frac{(1-p)^{\frac{\alpha-\beta}{\beta}}}
  {(1-p)^{\frac{\alpha-\beta}{\beta}}
  +p^{\frac{\alpha-\beta}{\beta}}}
  =
  \left[
    1+
    \left(\frac{p}{1-p}\right)^{\frac{\alpha-\beta}{\beta}}
  \right]^{-1}
  \in(0,1).
\]

Let
\[
  f(q)
  =
  q\left[
    1+
    \left(\frac{p}{1-p}\right)^{\frac{\alpha}{\beta}(\alpha-\beta)}
    \left(\frac{q}{1-q}\right)^{\alpha-1}
  \right].
\]
For $q\in(0,1)$,
\[
\begin{aligned}
  f'(q)
  &=
  1+
  \left(\frac{p}{1-p}\right)^{\frac{\alpha}{\beta}(\alpha-\beta)}
  \left(\frac{q}{1-q}\right)^{\alpha-1}
  \left[1+\frac{\alpha-1}{1-q}\right],\\
  f''(q)
  &=
  \alpha(\alpha-1)
  \left(\frac{p}{1-p}\right)^{\frac{\alpha}{\beta}(\alpha-\beta)}
  q^{\alpha-2}(1-q)^{-\alpha-1}.
\end{aligned}
\]
Thus, on $(0,q_{\alpha,\beta}(p))$, the function $v^{II}$ is strictly convex if $\alpha>1$, linear if $\alpha=1$, and strictly concave if $\alpha\in(0,1)$.

Since $p<q_{\alpha,\beta}(p)$ is equivalent to $p<1/2$, the value of the dynamic persuasion problem is
\[
  \left[\operatorname{CAV}(v^{II})\right](p)
  =
  \begin{cases}
    v^{II}(p)
    &\text{if }\alpha\in(0,1],\\
    \dfrac{p}{q_{\alpha,\beta}(p)}
    &\text{if }\alpha>1.
  \end{cases}
\]
Moreover,
\[
  v^{II}(p)
  =
  \begin{cases}
    p\left[
      1+
      \left(\frac{p}{1-p}\right)^{\frac{\alpha^2}{\beta}-1}
    \right]
    &\text{if }p<\frac{1}{2},\\
    1
    &\text{if }p\geq \frac{1}{2},
  \end{cases}
\]
and
\[
  \frac{p}{q_{\alpha,\beta}(p)}
  =
  p\left[
    1+
    \left(\frac{p}{1-p}\right)^{\frac{\alpha}{\beta}-1}
  \right].
\]
Hence, when $p\in(0,1/2)$,
\[
  \left[\operatorname{CAV}(v^{II})\right](p)
  =
  \begin{cases}
    p\left[
      1+
      \left(\frac{p}{1-p}\right)^{\frac{\alpha^2}{\beta}-1}
    \right]
    &\text{if }\alpha\in(0,1],\\
    p\left[
      1+
      \left(\frac{p}{1-p}\right)^{\frac{\alpha}{\beta}-1}
    \right]
    &\text{if }\alpha>1.
  \end{cases}
\]
This is the value of the dynamic persuasion problem.

Therefore, if $p<1/2$,
\[
  \sup_{x\in\mathcal{V}^2(p,D)}x
  -
  \sup_{x\in\mathcal{V}^1(p,D)}x
  =
  \begin{cases}
    p\left(\frac{p}{1-p}\right)^{\frac{\alpha}{\beta}-1}
    \left[
      \left(\frac{p}{1-p}\right)^{\frac{\alpha(\alpha-1)}{\beta}}
      -1
    \right]
    >0
    &\text{if }\alpha\in(0,1),\\
    0
    &\text{if }\alpha\geq 1.
  \end{cases}
\]
Thus,
\[
  \sup_{x\in\mathcal{V}^2(p,D)}x
  \begin{cases}
    >\sup_{x\in\mathcal{V}^1(p,D)}x
    &\text{if }\alpha\in(0,1),\\
    =\sup_{x\in\mathcal{V}^1(p,D)}x
    &\text{if }\alpha\geq 1.
  \end{cases}
\]
This completes the proof.


\subsection{Proof of Lemma \ref{lem:T1_equal_T2_iff_divisible}}

\textit{Only-if part}. By Lemma \ref{lem:divisible_then_T1_equal_T20}, if $D$ is divisible, then $T^1(p,D) = T^2_0(p,D)$ for every finite state space $\Theta$ and full-support prior $p$. By Lemma \ref{lem:T20_dense_in_T2}, $T^2_0(p,D)$ is dense in $T^2(p,D)$. Since $T^1(p,D)$ is closed by Lemma \ref{lem:T1_nonempty_compact_convex}, we have $T^1(p,D) = T^2(p,D)$.

\textit{If part}. If $T^1(p,D) = T^2(p,D)$ for every finite state space $\Theta$ and full-support prior $p$, then $T^1(p,D) \supseteq T^2_0(p,D)$ because $T^2_0(p,D)\subseteq T^2(p,D)$. Lemma \ref{lem:T20_sub_T1_then_divisible} then implies that $D$ is divisible.


\subsection{Proof of Theorem \ref{thm:divisible_iff_indifference}}\label{app:proof_thm_divisible_iff_indifference}

\subsubsection*{\(1\implies 2\)}

Fix any persuasion environment $(\Theta,p,A,u,v)$ and suppose the receiver's distortion rule $D$ is divisible. By Lemma \ref{lem:T1_equal_T2_iff_divisible}, divisibility implies
\[
  T^1(p,D)=T^2(p,D).
\]
Therefore, the feasible sets of ex ante sender payoffs induced by static and dynamic persuasion coincide:
\[
  \mathcal{V}^1(p,D)=\mathcal{V}^2(p,D).
\]

\subsubsection*{\(2\implies 3\)}

Immediate.

\subsubsection*{\(3\implies 1\)}

Fix any regular distortion rule $D$ satisfying prior independence under certainty, and suppose 3 holds. Consider any persuasion environment with $A=\Delta(\Theta)$ and
\[
  u(a,\theta)=-\|a-\delta_\theta\|^2.
\]
Then $\hat a(q')=q'$. If the sender's payoff is state-independent, so that $v(a,\theta)=v(a)$ for all $\theta\in\Theta$, then
\[
  \hat v(q,q')=v(q')
\]
for every sender posterior $q$. Thus, 3 implies that
\[
  \sup_{\tau\in \bar{T}^1(p,D)} \int_{\Delta(\Theta)} v\,d\tau
  =
  \sup_{\tau\in \bar{T}^2(p,D)} \int_{\Delta(\Theta)} v\,d\tau
\]
for every bounded continuous function $v:\Delta(\Theta)\to\mathbb R$.

We show that this equality requires $\bar{T}^2_0(p,D)\subseteq \bar{T}^1(p,D)$. Suppose, to the contrary, that there exists
\[
  \tau\in \bar{T}^2_0(p,D)\setminus \bar{T}^1(p,D).
\]
Since $\bar{T}^2_0(p,D)\subseteq \bar{T}^2(p,D)$, we also have $\tau\in \bar{T}^2(p,D)\setminus \bar{T}^1(p,D)$.

By Lemma \ref{lem:bar_T1_nonempty_compact_convex}, $\bar{T}^1(p,D)$ is a nonempty, compact, and convex subset of the space of regular signed measures of bounded variation on $\Delta(\Theta)$, denoted by $\mathcal{M}(\Delta(\Theta))$, endowed with the topology of weak convergence. Since $\mathcal{M}(\Delta(\Theta))$ is a locally convex topological vector space, Corollary 5.80 of \cite{aliprantis2006infinite} implies that there exists a nonzero continuous linear functional
\[
  f:\mathcal{M}(\Delta(\Theta))\to\mathbb R
\]
that strongly separates $\bar{T}^1(p,D)$ and $\tau$.

Let $C_b(\Delta(\Theta))$ be the set of bounded continuous functions from $\Delta(\Theta)$ to $\mathbb R$. The pair
\[
  \langle \mathcal{M}(\Delta(\Theta)), C_b(\Delta(\Theta)) \rangle
\]
is a dual pair under the bilinear form
\[
  (\tau,v)\mapsto \int_{\Delta(\Theta)} v\,d\tau.
\]
Thus, by Theorem 5.93 of \cite{aliprantis2006infinite}, every $\sigma(\mathcal{M}(\Delta(\Theta)),C_b(\Delta(\Theta)))$-continuous linear functional has the form
\[
  f(\tau)=\int_{\Delta(\Theta)} v(q')\,d\tau(q')
\]
for a unique $v\in C_b(\Delta(\Theta))$.

Since $\sigma(\mathcal{M}(\Delta(\Theta)),C_b(\Delta(\Theta)))$ is the topology of weak convergence on $\mathcal{M}(\Delta(\Theta))$, the separating functional can be represented by some $v\in C_b(\Delta(\Theta))$. Hence,
\[
  \sup_{\tau\in \bar{T}^1(p,D)} \int_{\Delta(\Theta)} v\,d\tau
  \neq
  \sup_{\tau\in \bar{T}^2(p,D)} \int_{\Delta(\Theta)} v\,d\tau,
\]
which contradicts 3. Therefore,
\[
  \bar{T}^2_0(p,D)\subseteq \bar{T}^1(p,D).
\]

By Lemma \ref{lem:bar_T20_sub_bar_T1_then_matrix_divisible}, for this fixed $p$, for every $q\in\Delta^\circ(\Theta)$, there exists a full-rank $|\Theta|\times|\Theta|$ column-stochastic matrix $M_{p,q}$ such that
\[
  D_p(M_{p,q}r)=D^{II}_q(r;p)
  \quad\text{for every } r\in\Delta(\Theta).
\]
Fix any $q\in\Delta^\circ(\Theta)$ and $\theta\in\Theta$. Setting $r=\delta_\theta$ gives
\[
  D_p(M_{p,q}\delta_\theta)
  =
  D^{II}_q(\delta_\theta;p)
  =
  D_{D_p(q)}(\delta_\theta).
\]
By prior independence under certainty,
\[
  D_{D_p(q)}(\delta_\theta)=D_p(\delta_\theta).
\]
Hence,
\[
  D_p(M_{p,q}\delta_\theta)=D_p(\delta_\theta)
\]
for every $\theta\in\Theta$. Since $D_p$ is one-to-one by regularity,
\[
  M_{p,q}\delta_\theta=\delta_\theta
\]
for every $\theta\in\Theta$. Thus, $M_{p,q}$ is the identity matrix. Therefore,
\[
  D_p(r)=D^{II}_q(r;p)
  \quad\text{for every } r\in\Delta(\Theta),
\]
which establishes divisibility.

\section{Functional Form of Divisible Updating Rules}\label{app:divisible_functional_form}

The next result makes it easier to check whether a candidate distortion rule is divisible. With one additional axiom, we obtain a clean characterization of divisible distortion rules, although we do not use it to prove our main results.

\begin{dfn}
  A regular distortion rule $D$ satisfies \textbf{approachability} if there exists a full-support belief $p^* \in \Delta^\circ(\Theta)$ such that the restriction $D_{p^*}\big|_{\Delta^\circ(\Theta)}:\Delta^\circ(\Theta)\to\Delta^\circ(\Theta)$ is surjective.
\end{dfn}
That is, approachability requires that, for some full-support prior, the restriction of the distortion rule to full-support beliefs be surjective onto the set of full-support beliefs. Under regularity and approachability, divisible distortion rules admit the following functional-form characterization.

\begin{prop}[cf. \cite{cripps2018divisible}]\label{prop:divisible_homeomorphism}
  Suppose a regular distortion rule $D$ satisfies approachability. Then $D$ is divisible if and only if there exists a homeomorphism $F: \Delta^\circ(\Theta) \to \Delta^\circ(\Theta)$ such that, for all full-support beliefs $p,q$,
  \begin{equation}\label{eqn:divisible_functional_form}
    D_p(q)=F^{-1}\left(\frac{\frac{F(p)}{p}q}{\left\langle \frac{F(p)}{p}, q \right\rangle}\right).
  \end{equation}
\end{prop}
This result is essentially a restatement of Proposition 1 in \cite{cripps2018divisible}. Because our statement differs slightly, we provide a proof. At a high level, the proof transforms divisibility into a so-called \textit{translation equation}, whose solutions are characterized by \cite{aczel1956transformations}.

Proposition \ref{prop:divisible_homeomorphism} immediately implies that geometric distortion is divisible, since it is a regular distortion rule satisfying approachability. Indeed, taking
\[
  F(q):=\frac{(q(\theta)^{1/\beta})_{\theta\in\Theta}}{\sum_{\theta'\in\Theta} q(\theta')^{1/\beta}}
\]
in \eqref{eqn:divisible_functional_form} yields the distortion rule for geometric distortion.

\subsection{Proof of Proposition \ref{prop:divisible_homeomorphism}}

The ``if'' part follows immediately by substituting \eqref{eqn:divisible_functional_form} into the definition of $D^{II}_q(r;p)$. We prove the ``only-if'' part.

Note that $D^{II}_q(r;p) = D_p(r)$ for all $p,q,r\in\Delta^\circ(\Theta)$ is equivalent to
\begin{equation}\label{eqn:proof_divisible_star_iff_invariance_1}
  D_{D_p(q)}\!\left(\frac{r\dfrac{D_p(q)}{q}}{\left\langle r, \dfrac{D_p(q)}{q} \right\rangle}\right)
  = D_p(r)
  \quad \text{for all } p,q,r\in\Delta^\circ(\Theta).
\end{equation}

Fix $\theta_0\in\Theta$. Define $u:\Delta^\circ(\Theta)\to\mathbb{R}^{|\Theta|-1}$ and, for each $x\in \Delta^\circ(\Theta)$, define $T_x: \mathbb{R}^{|\Theta|-1} \to \mathbb{R}^{|\Theta|-1}$ by
\[
  u(x)=\Bigl(\log\tfrac{x(\theta)}{x(\theta_0)}\Bigr)_{\theta\in\Theta\setminus\{\theta_0\}},
  \qquad
  T_x:=u\circ D_x\circ u^{-1}.
\]
Since $u$ is a homeomorphism, regularity of $D$ implies that $T_x$ is one-to-one for each $x$. Moreover, by approachability, there exists $p^*\in \Delta^\circ(\Theta)$ such that $T_{p^*}$ is onto. Define $H:\mathbb{R}^{|\Theta|-1}\times\mathbb{R}^{|\Theta|-1}\to \mathbb{R}^{|\Theta|-1}$ by
\[
  H(z,w):=T_{u^{-1}(z)}(w).
\]
Then $H(u(x),u(y))=T_x(u(y))$ for all $x,y\in\Delta^\circ(\Theta)$. Moreover, $H(z,\cdot)$ is one-to-one for every $z\in \mathbb{R}^{|\Theta|-1}$, and $H(u(p^*),\cdot)$ is onto.


Applying $u$ to \eqref{eqn:proof_divisible_star_iff_invariance_1} and using $u^{-1}\circ u=\mathrm{id}_{\Delta^\circ(\Theta)}$, we obtain
\[
  \begin{aligned}
  &T_{D_p(q)}\left[
    u\left(\frac{r \frac{D_p(q)}{q}}{\left\langle r,\frac{D_p(q)}{q} \right\rangle}\right)
  \right]
  = T_p(u(r)) \\
  \Leftrightarrow\quad
  &T_{D_p(q)}\left(u(r)+T_p(u(q))-u(q)\right)=T_p(u(r))\\
  \Leftrightarrow\quad
  &H\left(u(D_p(q)),u(r)+T_p(u(q))-u(q)\right)=H(u(p),u(r))\\
  \Leftrightarrow\quad
  &H\left(H(u(p),u(q)),u(r)+H(u(p),u(q))-u(q)\right)=H(u(p),u(r)).
  \end{aligned}
\]


  Rearranging the second argument of \(H\) and writing \(u(p),u(q),u(r)\) as \(u_p,u_q,u_r\in\mathbb{R}^{|\Theta|-1}\), we obtain
\begin{equation}\label{eqn:pre_translation}
  H\left(H(u_p,u_q), H(u_p,u_q)+u_r-u_q\right)=H(u_p,u_r).
\end{equation}

Define \(G:\mathbb{R}^{|\Theta|-1}\times\mathbb{R}^{|\Theta|-1}\to\mathbb{R}^{|\Theta|-1}\) by
\[
  G(\alpha,\beta):=H(\alpha,\alpha+\beta).
\]
Then
\[
  H(\alpha,\beta)=G(\alpha,\beta-\alpha).
\]
Substituting this into \eqref{eqn:pre_translation} gives
\[
  G\left(G(u_p,u_q-u_p),u_r-u_q\right)=G(u_p,u_r-u_p).
\]
Since \(u_r-u_p=(u_r-u_q)+(u_q-u_p)\), this is equivalent to
\[
  G\left(G(u_p,u_q-u_p),u_r-u_q\right)
  =
  G\left(u_p,(u_r-u_q)+(u_q-u_p)\right).
\]
Letting \(A:=u_q-u_p\) and \(B:=u_r-u_q\), we obtain
\begin{equation}\label{eqn:proof_divisible_star_iff_invariance_2}
  G(G(u_p,A),B)=G(u_p,A+B).
\end{equation}
 

  To solve this equation for $G$, we use the following corollary of Theorem 1 in \cite{aczel1956transformations}.

\begin{cor}\label{cor:translation_equation}
  Let $f:\mathbb{R}^{n}\times \mathbb{R}^n \to \mathbb{R}^{n}$. Suppose there exists $x^*\in \mathbb{R}^n$ such that $f(x^*,\cdot):\mathbb{R}^n\to\mathbb{R}^n$ is bijective. Then
  \[
    f(f(x,U),V)=f(x,U+V)
  \]
  for all $x,U,V\in\mathbb{R}^n$ if and only if there exists a bijection $g:\mathbb{R}^n\to\mathbb{R}^n$ such that
  \begin{equation}\label{eqn:sol_translation_equation}
    f(x,U)=g^{-1}\left(g(x)+U\right)
  \end{equation}
  for all $x,U\in\mathbb{R}^n$.
\end{cor}

\begin{proof}
  We first prove the ``if'' part. If \eqref{eqn:sol_translation_equation} holds for some bijection $g$, then
  \[
    \begin{aligned}
      f(f(x,U),V)
      &= g^{-1}\left(g(f(x,U))+V\right)\\
      &= g^{-1}\left(g\left(g^{-1}(g(x)+U)\right)+V\right)\\
      &= g^{-1}\left(g(x)+U+V\right)\\
      &= f(x,U+V).
    \end{aligned}
  \]

  We next prove the ``only-if'' part. Let $h:=f(x^*,\cdot):\mathbb{R}^n\to\mathbb{R}^n$. By assumption, $h$ is a bijection. Define $g:=h^{-1}$. Fix arbitrary $y,V\in\mathbb{R}^n$. Since $h$ is onto, there exists $U\in\mathbb{R}^n$ such that $y=h(U)=f(x^*,U)$; equivalently, $U=g(y)$. Using the translation equation with $x=x^*$, we obtain
  \[
    f(y,V)=f(f(x^*,U),V)=f(x^*,U+V)=h(U+V)=h(g(y)+V).
  \]
  Since $h=g^{-1}$ and $y,V$ are arbitrary,
  \[
    f(y,V)=g^{-1}\left(g(y)+V\right)
  \]
  for all $y,V\in\mathbb{R}^n$. Renaming $(y,V)$ as $(x,U)$ gives \eqref{eqn:sol_translation_equation}.
\end{proof}

  Recall that $H(u_x,\cdot)$ is one-to-one for any $u_x\in \mathbb{R}^{|\Theta|-1}$, and $H(u(p^*), \cdot)$ is onto. Since $H(\alpha,\beta) = G(\alpha, \beta-\alpha)$, $G(\alpha,\cdot)$ is also one-to-one for any $\alpha\in\mathbb{R}^{|\Theta|-1}$ and $G(u(p^*), \cdot)$ is onto. 


Therefore, by Corollary \ref{cor:translation_equation}, \(G\) satisfies \eqref{eqn:proof_divisible_star_iff_invariance_2} if and only if there exists a bijection \(g:\mathbb{R}^{|\Theta|-1}\to \mathbb{R}^{|\Theta|-1}\) such that
\[
  G(u_p,A)=g^{-1}\left(g(u_p)+A\right)
\]
for all \(u_p,A\in\mathbb{R}^{|\Theta|-1}\). Setting \(A=u_q-u_p\), we obtain
\[
  \begin{aligned}
    G(u_p,u_q-u_p)
    &= g^{-1}\left(g(u_p)+u_q-u_p\right)\\
    \Leftrightarrow\quad
    H(u_p,u_q)
    &= g^{-1}\left(g(u_p)+u_q-u_p\right)\\
    \Leftrightarrow\quad
    u(D_p(q))
    &= g^{-1}\left(g(u(p))+u(q)-u(p)\right)\\
    \Leftrightarrow\quad
    g(u(D_p(q)))
    &= g(u(p))+u(q)-u(p).
  \end{aligned}
\]

  Define $F:\Delta^\circ(\Theta)\to\Delta^\circ(\Theta)$ by
\[
  F:=u^{-1}\circ g\circ u.
\]
Then $F$ is a bijection and $u\circ F=g\circ u$. Hence,
\[
  g(u(D_p(q)))=u(F(p))+u(q)-u(p).
\]
By the definition of $u$, this is equivalent to
\[
  g(u(D_p(q)))
  =
  u\left(
    \frac{\frac{F(p)}{p}q}
    {\left\langle \frac{F(p)}{p},q\right\rangle}
  \right).
\]
Applying $u^{-1}$ to both sides gives
\[
  F(D_p(q))
  =
  \frac{\frac{F(p)}{p}q}
  {\left\langle \frac{F(p)}{p},q\right\rangle}.
\]
Applying \(F^{-1}\) to both sides yields
\[
  D_p(q)
  =
  F^{-1}\left(
  \frac{\frac{F(p)}{p}q}
  {\left\langle \frac{F(p)}{p},q\right\rangle}
  \right),
\]
for all $p,q\in\Delta^\circ(\Theta)$.


  It remains to show that $F:\Delta^\circ(\Theta)\to\Delta^\circ(\Theta)$ is a homeomorphism. For each $p\in\Delta^\circ(\Theta)$, define
\[
  B_p(q)
  :=
  \frac{\frac{F(p)}{p}q}
  {\left\langle \frac{F(p)}{p},q\right\rangle}.
\]
The map $B_p:\Delta^\circ(\Theta)\to\Delta^\circ(\Theta)$ is a homeomorphism. From the preceding argument, we have
\[
  F\circ D_p = B_p
  \quad \text{for all } p\in\Delta^\circ(\Theta).
\]

By approachability, there exists $p^*\in\Delta^\circ(\Theta)$ such that
\[
  D_{p^*}\big|_{\Delta^\circ(\Theta)}:\Delta^\circ(\Theta)\to\Delta^\circ(\Theta)
\]
is surjective. By regularity, $D_{p^*}$ is continuous and one-to-one. Hence $D_{p^*}$ is a continuous bijection from $\Delta^\circ(\Theta)$ onto itself.

To see that $D_{p^*}$ is a homeomorphism, identify $\Delta^\circ(\Theta)$ with the open set
\[
  U:=\left\{(x_1,\ldots,x_{n-1})\in\mathbb{R}^{n-1}_{++}:
  \sum_{k=1}^{n-1}x_k<1\right\},
  \qquad n:=|\Theta|,
\]
through the coordinate projection
\[
  \phi(p)=(p(\theta_1),\ldots,p(\theta_{n-1})).
\]
Then
\[
  \phi\circ D_{p^*}\circ \phi^{-1}:U\to U
\]
is a continuous bijection. By Brouwer's invariance of domain theorem, it is a homeomorphism. Therefore, $D_{p^*}$ is a homeomorphism.

Since
\[
  F\circ D_{p^*}=B_{p^*},
\]
we have
\[
  F = B_{p^*}\circ D_{p^*}^{-1}.
\]
Thus, $F$ is a homeomorphism as a composition of homeomorphisms. This completes the proof.

  \subsection{Proof of Theorem \ref{thm:weakly_divisibile_iff_indifference}}

\subsubsection*{\(1\implies 2\)}

Fix any persuasion environment $(\Theta,p,A,u,v)$ with state-independent sender preferences, and suppose the receiver has a regular distortion rule $D$ satisfying invariance under no information. 

If $D$ is weakly divisible, Lemma \ref{lem:weakly_divisible_to_bar_T20_sub_bar_T1} implies
\[
  \bar T^1(p,D)=\bar T^2(p,D).
\]
Since the sender's indirect utility depends only on the receiver's beliefs, the feasible sets of ex ante sender payoffs induced by static and dynamic persuasion coincide.

\subsubsection*{\(2\implies 3\)}

Immediate.

\subsubsection*{\(3\implies 1\)}

Suppose 3 holds. By the same separation argument as in the proof of \(3\implies 1\) in Theorem \ref{thm:divisible_iff_indifference}, applied to environments with state-independent sender preferences, we obtain
\[
  \bar T_0^2(p,D)\subseteq \bar T^1(p,D)
\]
for every finite state space $\Theta$ and full-support prior $p\in\Delta^\circ(\Theta)$.

By Lemma \ref{lem:bar_T20_sub_bar_T1_then_matrix_divisible}, for every $p,q\in\Delta^\circ(\Theta)$, there exists a full-rank $|\Theta|\times|\Theta|$ column-stochastic matrix $M_{p,q}$ such that
\[
  D_p(M_{p,q}r)=D_q^{II}(r;p)
  \quad\text{for every } r\in\Delta(\Theta).
\]
Since $D$ satisfies invariance under no information by assumption, we also have
\[
  D_p(p)=p
  \quad\text{for every } p\in\Delta^\circ(\Theta).
\]
Thus, $D$ is weakly divisible.

\section{Linear Underreaction Is Not Weakly Divisible}\label{app:linear_underreaction}

\begin{cor}\label{cor:linear_underreaction}
  Suppose the receiver's distortion rule is linear underreaction with $\lambda \in (0,1)$. Then there exists a persuasion environment in which the sender strictly prefers dynamic persuasion to static persuasion.
\end{cor}

\begin{proof}
  Suppose the receiver's distortion rule $D$ is given by
  \[
    D_p(q)=\lambda q+(1-\lambda)p
  \]
  for some $\lambda\in(0,1)$, for any finite state space $\Theta$, prior $p\in\Delta(\Theta)$, and belief $q\in\Delta(\Theta_p)$. It is enough to show that $D$ is not weakly divisible.

  Take $\Theta=\{0,1\}$ and identify each belief with its mass on state $1$. Fix $p=1/2$ and $q=1/4$. Then
  \[
    D_p(q)=\frac{2-\lambda}{4}.
  \]
  For any $r\in[0,1]$, the interim distortion is
  \[
    D^{II}_q(r;p)
    =
    \lambda
    \frac{3(2-\lambda)r}{4(1-\lambda)r+2+\lambda}
    +(1-\lambda)\frac{2-\lambda}{4}.
  \]
  This function is not affine in $r$ for any $\lambda\in(0,1)$.

  Now fix any column-stochastic matrix $M_{p,q}$. Then, for some $a,b\in[0,1]$,
  \[
    M_{p,q}
    =
    \begin{bmatrix}
      a & b\\
      1-a & 1-b
    \end{bmatrix}.
  \]
  Hence, identifying beliefs with their mass on state $1$,
  \[
    M_{p,q}r=ar+b(1-r).
  \]
  Therefore,
  \[
    D_p(M_{p,q}r)
    =
    \lambda\bigl(ar+b(1-r)\bigr)+\frac{1-\lambda}{2}
    =
    \lambda(a-b)r+\lambda b+\frac{1-\lambda}{2},
  \]
  which is affine in $r$. Since $D^{II}_q(r;p)$ is not affine in $r$, no column-stochastic matrix $M_{p,q}$ can satisfy
  \[
    D_p(M_{p,q}r)=D^{II}_q(r;p)
  \]
  for all $r\in[0,1]$. Thus, $D$ is not weakly divisible.

  Linear underreaction satisfies invariance under no information. Hence, by Theorem \ref{thm:weakly_divisibile_iff_indifference}, the sender is not indifferent between static and dynamic persuasion in some persuasion environment. Since invariance under no information implies that dynamic persuasion weakly dominates static persuasion, the sender strictly prefers dynamic persuasion in some environment.
\end{proof}
  \bibliography{ref}

@techreport{bohren2023behavioral,
  author      = {Bohren, J. Aislinn and Hauser, Daniel N.},
  title       = {The Behavioral Foundations of Model Misspecification: A Decomposition},
  institution = {Penn Institute for Economic Research, Department of Economics, University of Pennsylvania},
  year        = {2023}
}

@article{kamenica2011bayesian,
  title={Bayesian persuasion},
  author={Kamenica, Emir and Gentzkow, Matthew},
  journal={American Economic Review},
  volume={101},
  number={6},
  pages={2590--2615},
  year={2011},
  publisher={American Economic Association}
}

@article{de2022non,
  title={Non-bayesian persuasion},
  author={De Clippel, Geoffroy and Zhang, Xu},
  journal={Journal of Political Economy},
  volume={130},
  number={10},
  pages={2594--2642},
  year={2022},
  publisher={The University of Chicago Press Chicago, IL}
}

@article{cripps2018divisible,
  title={Divisible updating},
  author={Cripps, Martin W},
  journal={Manuscript, UCL},
  year={2018}
}

@article{che2023keeping,
  title={Keeping the listener engaged: a dynamic model of bayesian persuasion},
  author={Che, Yeon-Koo and Kim, Kyungmin and Mierendorff, Konrad},
  journal={Journal of Political Economy},
  volume={131},
  number={7},
  pages={1797--1844},
  year={2023},
  publisher={The University of Chicago Press Chicago, IL}
}

@article{rayo2010optimal,
  title={Optimal information disclosure},
  author={Rayo, Luis and Segal, Ilya},
  journal={Journal of political Economy},
  volume={118},
  number={5},
  pages={949--987},
  year={2010},
  publisher={University of Chicago Press Chicago, IL}
}

@article{alonso2016bayesian,
  title={Bayesian persuasion with heterogeneous priors},
  author={Alonso, Ricardo and C{\^a}mara, Odilon},
  journal={Journal of Economic Theory},
  volume={165},
  pages={672--706},
  year={2016},
  publisher={Elsevier}
}

@article{benjamin2019errors,
  title={Errors in probabilistic reasoning and judgment biases},
  author={Benjamin, Daniel J},
  journal={Handbook of Behavioral Economics: Applications and Foundations 1},
  volume={2},
  pages={69--186},
  year={2019},
  publisher={Elsevier}
}

@article{grether1980bayes,
  title={Bayes rule as a descriptive model: The representativeness heuristic},
  author={Grether, David M},
  journal={The Quarterly journal of economics},
  volume={95},
  number={3},
  pages={537--557},
  year={1980},
  publisher={MIT Press}
}

@article{aczel1956transformations,
  title={On transformations with several parameters and operations in multidimensional spaces},
  author={Acz{\'e}l, J and Hossz{\'u}, M},
  journal={Acta Mathematica Hungarica},
  volume={7},
  number={3-4},
  pages={327--338},
  year={1956},
  publisher={Akad{\'e}miai Kiad{\'o}, co-published with Springer Science+ Business Media BV~…}
}

@book{aliprantis2006infinite,
  title={Infinite dimensional analysis: a hitchhiker’s guide},
  author={Aliprantis, Charalambos D and Border, Kim C},
  year={2006},
  publisher={Springer}
}

@article{ba2025over,
  title={Over-and Underreaction to Information: Belief Updating with Cognitive Constraints},
  author={Ba, Cuimin and Bohren, J Aislinn and Imas, Alex},
  year={2025}
}

@article{strzalecki2024variational,
  title={Variational bayes and non-bayesian updating},
  author={Strzalecki, Tomasz},
  journal={arXiv preprint arXiv:2405.08796},
  year={2024}
}

@article{azrieli2025sequential,
  title={Sequential Non-Bayesian Persuasion},
  author={Azrieli, Yaron and Das, Rachana},
  journal={arXiv preprint arXiv:2508.09464},
  year={2025}
}

@article{bhuller2025feedback,
  title={Feedback and learning: The causal effects of reversals on judicial decision-making},
  author={Bhuller, Manudeep and Sigstad, Henrik},
  journal={Review of Economic Studies},
  volume={92},
  number={4},
  pages={2359--2397},
  year={2025},
  publisher={Oxford University Press UK}
}

@article{bordalo2020overreaction,
  title={Overreaction in macroeconomic expectations},
  author={Bordalo, Pedro and Gennaioli, Nicola and Ma, Yueran and Shleifer, Andrei},
  journal={American Economic Review},
  volume={110},
  number={9},
  pages={2748--2782},
  year={2020},
  publisher={American Economic Association 2014 Broadway, Suite 305, Nashville, TN 37203}
}

@article{levy2022persuasion,
  title={Persuasion with correlation neglect: a full manipulation result},
  author={Levy, Gilat and Barreda, In{\'e}s Moreno de and Razin, Ronny},
  journal={American Economic Review: Insights},
  volume={4},
  number={1},
  pages={123--138},
  year={2022},
  publisher={American Economic Association 2014 Broadway, Suite 305, Nashville, TN 37203}
}

@article{epstein2010non,
  title={Non-bayesian learning},
  author={Epstein, Larry G and Noor, Jawwad and Sandroni, Alvaro and others},
  journal={The BE Journal of Theoretical Economics},
  volume={10},
  number={1},
  pages={1--20},
  year={2010},
  publisher={De Gruyter}
}

@article{esponda2016berk,
  title={Berk--Nash equilibrium: A framework for modeling agents with misspecified models},
  author={Esponda, Ignacio and Pouzo, Demian},
  journal={Econometrica},
  volume={84},
  number={3},
  pages={1093--1130},
  year={2016},
  publisher={Wiley Online Library}
}

@article{yang2026stochastic,
  title={Stochastic Optimization and Coupling},
  author={Yang, Frank and Yang, Kai Hao},
  journal={arXiv preprint arXiv:2603.11448},
  year={2026}
}

@article{chambers2023coherent,
  title={Coherent distorted beliefs},
  author={Chambers, Christopher P and Masatlioglu, Yusufcan and Raymond, Collin},
  journal={arXiv preprint arXiv:2310.09879},
  year={2023}
}

\end{document}